

\message{Preloading the phys format:}


\catcode`\@=11

\chardef\f@ur=4
\chardef\l@tter=11
\chardef\@ther=12
\dimendef\dimen@iv=4
\toksdef\toks@i=1 
\toksdef\toks@ii=2
\newtoks\emptyt@ks 

\def\glet{\global\let}
\def\gz@#1{\global#1\z@}
\def\gm@ne#1{\global#1\m@ne}
\def\g@ne#1{\global\advance#1\@ne}

\def\@height{height}       
\def\@depth{depth}         
\def\@width{width}         

\def\@plus{plus}           
\def\@minus{minus}         

\message{macros for text,}


\def\loop#1\repeat{\def\iter@te{#1\expandafter\iter@te \fi}\iter@te
  \let\iter@te\undefined}

\def\bpargroup{\bp@rgroup\ep@r} 
\def\bgrafgroup{\bp@rgroup\ep@rgroup} 
\def\bp@rgroup{\bgroup \let\par\ep@rgroup \let\endgraf}

\def\ep@r{\ifhmode \unpenalty\unskip \fi \p@r}
\def\ep@rgroup{\ep@r \egroup}

\let\p@r=\endgraf   
\let\par=\ep@r
\let\endgraf=\ep@r

\def\lb{\hfil\break}
\def\endpage{\par \vfil \eject}
\def\superendpage{\par \vfil \supereject}



\def\leftline{\@line\hsize\empty\hss}
\def\rightline{\@line\hsize\hss\empty}
\def\centerline{\@line\hsize\hss\hss}

\let\plainrlap=\rlap   
\let\plainllap=\llap   

\def\rlap{\@line\z@\empty\hss}
\def\llap{\@line\z@\hss\empty}

\def\lftline{\@line\hsize\empty\hfil}
\def\rtline{\@line\hsize\hfil\empty}
\def\ctrline{\@line\hsize\hfil\hfil}

\def\@line#1#2#3{\hbox to#1\bgroup#2\let\n@xt#3%
  \afterassignment\@@line \setbox\z@\hbox}
\def\@@line{\aftergroup\@@@line}
\def\@@@line{\unhbox\z@ \n@xt\egroup}

\def\after@arg#1{\bgroup\aftergroup#1\afterassignment\after@@arg\@eat}
\def\after@@arg{\ifcat\bgroup\noexpand\n@xt\else \n@xt\egroup \fi}
\def\@eat{\let\n@xt= } 
\def\eat#1{}           
\def\@eat@#1{\@eat}    


\let\nl=\space

\def\ctrlines#1#2{\par \bpargroup
  \bgroup \parskip\z@skip \noindent \egroup
  \let\ctr@style#1\let\nl\ctr@lines \hfil \ctr@style{#2}\strut
  \interlinepenalty\@M \par}
\def\ctr@lines{\strut \lb \strut \hfil \ctr@style}


\def\begin@{\ifmmode \expandafter\mathpalette\expandafter\math@ \else
  \expandafter\make@ \fi}
\def\make@#1{\setbox\z@\hbox{#1}\fin@}
\def\math@#1#2{\setbox\z@\hbox{$\m@th#1{#2}$}\fin@}

\def\ph@nt{\let\fin@\finph@nt \begin@}
\let\makeph@nt=\undefined
\let\mathph@nt=\undefined

\newif\ift@ \newif\ifb@
\def\topsmash{\t@true\b@false\sm@sh}
\def\botsmash{\t@false\b@true\sm@sh}
\def\smash{\t@true\b@true\sm@sh}
\def\sm@sh{\let\fin@\finsm@sh \begin@}
\let\makesm@sh=\undefined
\let\mathsm@sh=\undefined
\def\finsm@sh{\ift@\ht\z@\z@\fi \ifb@\dp\z@\z@\fi \box\z@}


\newdimen\boxitsep   \boxitsep=5pt

\def\fboxit#1#2{\hbox{\vrule \@width#1\p@
    \vtop{\vbox{\hrule \@height#1\p@ \vskip\boxitsep
        \hbox{\hskip\boxitsep #2\hskip\boxitsep}}%
      \vskip\boxitsep \hrule \@height#1\p@}\vrule \@width#1\p@}}


\begingroup
  \catcode`\:=\active
  \lccode`\*=`\\ \lowercase{\gdef:{*}}   
  \catcode`\;=\active
  \lccode`\* `\% \lowercase{\gdef;{*}}   
  \catcode`\^^M=\active \glet^^M=\space  
\endgroup


\begingroup
  \catcode`\:=\active
  \outer\gdef\comment{\begingroup
    \catcode`\\\@ther \catcode`\%\@ther \catcode`\^^M\@ther
    \catcode`\{\@ther \catcode`\}\@ther \catcode`\#\@ther
    \wlog{* input between `:comment' and `:endcomment' ignored *}%
    \c@mment}
\endgroup
{\lccode`\:=`\\ \lccode`\;=`\^^M
  \lowercase{\gdef\c@mment#1;{\c@@mment:endcomment*}}}
\def\c@@mment#1#2*#3{\if #1#3%
    \ifx @#2@\def\n@xt{\endgroup\ignorespaces}\else
      \def\n@xt{\c@@mment#2*}\fi \else
    \def\n@xt{\c@mment#3}\fi \n@xt}

\message{date and time,}


\newcount\langu@ge

\let\mainlanguage\relax

\begingroup \catcode`\"=\@ther \gdef\dq{"}
  \catcode`\"=\active
  \gdef"#1{\ifx#1s\ss\else\ifx#1SSS\else
    {\accent\dq 7F #1}\penalty\@M \hskip\z@skip \fi\fi}
  \endgroup

\outer\def\english{\gm@ne\langu@ge
  \global\catcode`\"\@ther \glet\3\undefined}
\outer\def\german{\gz@\langu@ge
  \global\catcode`\"\active \glet\3\ss}

\def\case@language#1{\ifcase\expandafter\langu@ge #1\fi}
\def\case@abbr#1{{\let\nodot\n@dot\case@language{#1}.~}}
\def\n@dot{\expandafter\eat}
\let\nodot=\empty


\def\themonth{\xdef\themonth{\noexpand\case@language
  {\ifcase\month \or Januar\or Februar\or M\noexpand\"arz\or April\or
  Mai\or Juni\or Juli\or August\or September\or Oktober\or November\or
  Dezember\fi
  \noexpand\else
  \ifcase\month \or January\or February\or March\or April\or May\or
  June\or July\or August\or September\or October\or November\or
  December\fi}}\themonth}

\def\thedate{\case@language{\else\themonth\ }\number\day
  \case@language{.\ \themonth \else ,} \number\year}

\def\date{\number\day.\,\number\month.\,\number\year}


\def\PhysTeX{$\Phi\kern-.25em\raise.4ex\hbox{$\Upsilon$}\kern-.225em
  \Sigma$-\TeX}

\message{the time,}



\count255=\time \divide\count255 by 60 \edef\thetime{\the\count255 :}
\multiply\count255 by -60 \advance\count255 by\time
\edef\thetime{\thetime \ifnum10>\count255 0\fi \the\count255 }

\message{spacing, fonts and sizes,}


\newskip\refbetweenskip   \newskip\chskiptamount
\newskip\chskiplamount   \newskip\secskipamount
\newskip\footnotebaselineskip   \newskip\interfootnoteskip

\newdimen\chapstretch   \chapstretch=2.5cm
\newcount\chappenalty   \chappenalty=-800
\newdimen\sectstretch   \sectstretch=2cm
\newcount\sectpenalty   \sectpenalty=-400

\def\chskipt{\chapbreak \vskip\chskiptamount}
\def\chskipl{\nobreak \vskip\chskiplamount}
\def\unchskip{\vskip-\chskiplamount}
\def\secskipt{\sectbreak \vskip\secskipamount}
\def\chapbreak{\par \vskip\z@\@plus\chapstretch \penalty\chappenalty
  \vskip\z@\@plus-\chapstretch}
\def\sectbreak{\par \vskip\z@\@plus\sectstretch \penalty\sectpenalty
  \vskip\z@\@plus-\sectstretch}



\begingroup \lccode`\*=`\r
  \lowercase{\def\n@xt#1*#2@{#1}
    \xdef\font@sel{\expandafter\n@xt\fontname\tenrm*@}}\endgroup

\message{loading \font@sel\space fonts,}

=\font@sel r12 
=\font@sel r9
=\font@sel r8
=\font@sel r6

=\font@sel mi12 \skewchar\twelvei='177 
=\font@sel mi9    \skewchar\ninei='177
=\font@sel mi8   \skewchar\eighti='177
=\font@sel mi6   \skewchar\sixi='177

=\font@sel sy10 scaled \magstep1
  \skewchar\twelvesy='60 
=\font@sel sy9    \skewchar\ninesy='60
=\font@sel sy8   \skewchar\eightsy='60
=\font@sel sy6   \skewchar\sixsy='60




=\font@sel bx12 
=\font@sel bx9
=\font@sel bx8
=\font@sel bx6

=\font@sel tt12 
=\font@sel tt8


=\font@sel sl12 
=\font@sel sl9
=\font@sel sl8

\font\twelveit=\font@sel ti12 
\font\nineit=\font@sel ti9
\font\eightit=\font@sel ti8
=\font@sel ti7



=\font@sel csc10 scaled \magstep1 
=\font@sel csc10






\def\twelvepoint{\twelve@point
  \let\normal@spacing\twelve@spacing \set@spacing}
\def\tenpoint{\ten@point
  \let\normal@spacing\ten@spacing \set@spacing}
\def\eightpoint{\eight@point
  \let\normal@spacing\eight@spacing \set@spacing}

\def\rm{\fam\z@ \@fam}
\def\mit{\fam\@ne}
\def\oldstyle{\mit \@fam}
\def\cal{\fam\tw@}
\def\it{\fam\itfam \@fam}
\def\sl{\fam\slfam \@fam}
\def\bf{\fam\bffam \@fam}
\def\tt{\fam\ttfam \@fam}
\def\caps{\@caps}
\def\@fam{\the\textfont\fam}


\def\twelve@point{\set@fonts twelve ten eight }
\def\ten@point{\set@fonts ten eight six }
\def\eight@point{\set@fonts eight six five }

\def\set@fonts#1 #2 #3 {\textfont\ttfam\csname#1tt\endcsname
    \expandafter\let\expandafter\@caps\csname#1csc\endcsname
  \def\n@xt##1##2{\textfont##1\csname#1##2\endcsname
    \scriptfont##1\csname#2##2\endcsname
    \scriptscriptfont##1\csname#3##2\endcsname}%
  \set@@fonts}
\def\set@@fonts{\n@xt0{rm}\n@xt1i\n@xt2{sy}%
  \textfont3\tenex \scriptfont3\tenex \scriptscriptfont3\tenex
  \n@xt\itfam{it}\n@xt\slfam{sl}\n@xt\bffam{bf}\rm}


\def\singlespace{\chardef\@spacing\z@ \set@spacing}
\def\doublespace{\chardef\@spacing\@ne \set@spacing}
\def\triplespace{\chardef\@spacing\tw@ \set@spacing}
\chardef\@spacing=1   

\def\set@spacing{\expandafter\expandafter\expandafter\set@@spacing
  \expandafter\spacing@names\expandafter\@@\normal@spacing
  \normalbaselines}
\def\set@@spacing#1#2\@@#3+#4*{#1#4\multiply#1\@spacing \advance#1#3%
  \ifx @#2@\let\n@xt\empty \else
    \def\n@xt{\set@@spacing#2\@@}\fi \n@xt}

\def\normalbaselines{\lineskip\normallineskip
  \setbaselineskip\normalbaselineskip
  \lineskiplimit\normallineskiplimit}

\def\setbaselineskip{\afterassignment\set@strut \baselineskip}
\def\set@strut{\setbox\strutbox\spacer\z@\baselineskip}

\def\spacer{\hbox\bgroup \afterassignment\x@spacer \dimen@}
\def\x@spacer{\ifdim\dimen@=\z@\else \hskip\dimen@ \fi
  \afterassignment\y@spacer \dimen@}
\def\y@spacer{\setbox\z@\hbox{$\vcenter{\vskip\dimen@}$}%
  \vrule \@height\ht\z@ \@depth\dp\z@ \@width\z@ \egroup}

\def\spacing@names{
  \normalbaselineskip
  \normallineskip
  \normallineskiplimit
  \footnotebaselineskip
  \interfootnoteskip
  \parskip
  \refbetweenskip
  \abovedisplayskip
  \belowdisplayskip
  \abovedisplayshortskip
  \belowdisplayshortskip
  \chskiptamount
  \chskiplamount
  \secskipamount
  }

\def\twelve@spacing{
  14\p@              +5\p@        *
  \p@                +\z@         *
  \z@                +\z@         *
  14\p@              +\p@         *
  20\p@              +\z@         *
  5\p@\@plus\p@      +-2\p@       *
  \z@                +4\p@        *
  8\p@\@plus2\p@\@minus3\p@  +%
    4\p@\@plus3\p@\@minus5\p@     *
  8\p@\@plus2\p@\@minus3\p@  +%
    4\p@\@plus3\p@\@minus5\p@     *
  \p@\@plus2\p@\@minus\p@    +%
    4\p@\@plus3\p@\@minus2\p@     *
  8\p@\@plus2\p@\@minus3\p@  +%
    \p@\@plus2\p@\@minus2\p@      *
  20\p@\@plus5\p@    +\z@         *
  5.5\p@             +\z@         *
  6\p@\@plus2\p@     +\z@         *
  }

\def\ten@spacing{
  11\p@              +4.5\p@      *
  \p@                +\z@         *
  \z@                +\z@         *
  12\p@              +\p@         *
  16\p@              +\z@         *
  5\p@\@plus\p@      +-2\p@       *
  \z@                +5\p@        *
  8\p@\@plus2\p@\@minus3\p@  +%
    4\p@\@plus3\p@\@minus5\p@     *
  8\p@\@plus2\p@\@minus3\p@  +%
    4\p@\@plus3\p@\@minus5\p@     *
  \p@\@plus2\p@\@minus\p@    +%
    4\p@\@plus3\p@\@minus2\p@     *
  8\p@\@plus2\p@\@minus3\p@  +%
    \p@\@plus2\p@\@minus2\p@      *
  20\p@\@plus5\p@    +\z@         *
  5.5\p@             +\z@         *
  6\p@\@plus2\p@     +\z@         *
  }

\def\eight@spacing{
  9\p@               +3.5\p@      *
  \p@                +\z@         *
  \z@                +\z@         *
  10\p@              +\p@         *
  14\p@              +\z@         *
  5\p@\@plus\p@      +-2\p@       *
  \z@                +5\p@        *
  8\p@\@plus2\p@\@minus3\p@  +%
    4\p@\@plus3\p@\@minus5\p@     *
  8\p@\@plus2\p@\@minus3\p@  +%
    4\p@\@plus3\p@\@minus5\p@     *
  \p@\@plus2\p@\@minus\p@    +%
    4\p@\@plus3\p@\@minus2\p@     *
  8\p@\@plus2\p@\@minus3\p@  +%
    \p@\@plus2\p@\@minus2\p@      *
  20\p@\@plus5\p@    +\z@         *
  5.5\p@             +\z@         *
  6\p@\@plus2\p@     +\z@         *
  }

\twelvepoint


\def\large{\par \bgroup \twelvepoint \after@arg\@size}
\def\medium{\par \bgroup \tenpoint \after@arg\@size}
\def\small{\par \bgroup \eightpoint \after@arg\@size}
\def\@size{\par \egroup}

\def\LARGE#1{{\twelve@point #1}}
\def\MEDIUM#1{{\ten@point #1}}
\def\SMALL#1{{\eight@point #1}}

\message{texts, headings and styles,}


\def\submittextone{Zur Ver\"offentlichung in\else Submitted to}
\def\submittexttwo{ eingereicht\else}
\def\abstracthead{Zusammenfassung\else Abstract}
\def\ackhead{Danksagung\else Acknowledgements}
\def\appendixhead{Anhang\else Appendix}
\def\eqabbr{Gl\else eq}
\def\eqsabbr{Gln\else eqs}
\def\figpref{Abb\else Fig}
\def\fighead{Abbildungen\else Figure captions}
\def\figabbr{Bild\nodot\else Fig}
\def\figsabbr{Bilder\nodot\else Figs}
\def\tabpref{Tab\else Tab}
\def\tabhead{Tabellen\else Table captions}
\let\tababbr=\tabpref
\def\tabsabbr{Tab\else Tabs}
\def\refpref{Lit\else Ref}
\def\refhead{Literaturverzeichnis\else References}
\def\refabbr{???\else Ref}
\def\refsabbr{????\else Refs}
\def\tocpref{Inh\else Toc}
\def\tochead{Inhaltsverzeichnis\else Table of contents}
\def\footpref{Anm\else Foot}
\def\foothead{Anmerkungen\else Footnotes}
\def\prfhead{Beweis\else Proof}


\def\UPPERCASE#1{\edef\n@xt{#1}\uppercase\expandafter{\n@xt}}

\let\headlinestyle=\twelverm
\let\footlinestyle=\twelverm
\let\pagestyle=\twelverm       
\let\titlestyle=\bf            
\let\authorstyle=\caps
\let\addressstyle=\sl
\let\sectstyle=\caps           

\let\headstyle=\UPPERCASE      
\let\captionstyle=\it          
\let\journalstyle=\sl
\let\volumestyle=\bf



\def\namrefindent{2em}


\let\footstyle=\empty          

\let\stmttitlestyle=\bf        
\let\stmtstyle=\sl
\let\prftitlestyle=\caps
\let\prfstyle=\sl


\def\skipuserexit{\setbox\z@\box\@cclv}  
\def\shipuserexit{\unvbox\@cclv}         

\def\chapuserexit{\sectuserexit}         
\def\appuserexit{\chapuserexit}          
\def\sectuserexit{\secsuserexit}         
\let\secsuserexit=\relax                 

\message{page numbers and output,}


\newcount\firstp@ge   \firstp@ge=-10000
\newcount\lastp@ge   \lastp@ge=10000
\outer\def\pagesel#1#2{\global\firstp@ge#1 \global\lastp@ge#2
  \wlog{**************************}\wlog{*}%
  \wlog{* don't use \string\p agesel}%
  \wlog{* this will not be supported in future}%
  \wlog{*}\wlog{**************************}%
  \wlog{(* pages #1-#2 selected for printing, others will be skipped *)}}

\newbox\pageb@x

\outer\def\toppagenum{\glet\page@tbn T%
  \glet\headb@x\pageb@x \glet\footb@x\voidb@x}
\outer\def\botpagenum{\glet\page@tbn B%
  \glet\headb@x\voidb@x \glet\footb@x\pageb@x}
\outer\def\nopagenum{\glet\page@tbn N%
  \glet\headb@x\voidb@x \glet\footb@x\voidb@x}

\outer\def\lefthead{\glet\head@lrac L}
\outer\def\righthead{\glet\head@lrac R}
\outer\def\althead{\glet\head@lrac A}
\outer\def\centhead{\glet\head@lrac C}
\outer\def\leftfoot{\glet\foot@lrac L}
\outer\def\rightfoot{\glet\foot@lrac R}
\outer\def\altfoot{\glet\foot@lrac A}
\outer\def\centfoot{\glet\foot@lrac C}

\newtoks\lheadtext   \newtoks\cheadtext   \newtoks\rheadtext
\newtoks\lfoottext   \newtoks\cfoottext   \newtoks\rfoottext

\headline={\headlinestyle \head@foot\skip@head\head@lrac
  \lheadtext\cheadtext\rheadtext\headb@x}
\footline={\footlinestyle \head@foot\skip@foot\foot@lrac
  \lfoottext\cfoottext\rfoottext\footb@x}
\lheadtext={}   \cheadtext={}   \rheadtext={}
\lfoottext={}   \cfoottext={}   \rfoottext={}

\newbox\page@strut
\setbox\page@strut\hbox{\vrule \@height 15mm\@depth 10mm\@width \z@}

\def\head@foot#1#2#3#4#5#6{\unhcopy\page@strut
  \if#1T\hfil \else
    \if#2C\head@@foot{\the#3}{\copy#6}{\the#5}\else
      \if#2A\ifodd\pageno \let#2R\else \let#2L\fi \fi
      \if#2R\head@@foot{\the#4}{\the#5}{\copy#6}\else
        \head@@foot{\copy#6}{\the#3}{\the#4}\fi \fi \fi}
\def\head@@foot#1#2#3{\plainrlap{#1}\hfil#2\hfil\plainllap{#3}}

\let\startpage=\relax  

\outer\def\pageall{\glet\page@ac A%
  \global\countdef\pageno\z@ \global\pageno\@ne
  \global\countdef\pageno@pref\@ne \gz@\pageno@pref
  \glet\page@pref\empty \glet\page@reset\count@
  \glet\chap@break\chskipt \outer\gdef\startpage{\global\pageno}}
\outer\def\pagechap{\glet\page@ac C%
  \global\countdef\pageno\@ne \gz@\pageno
  \global\countdef\pageno@pref\z@ \gz@\pageno@pref
  \gdef\page@pref{\dash@pref}%
  \gdef\page@reset{\global\pageno\@ne \global\pageno@pref}%
  \glet\chap@break\superendpage
  \outer\gdef\startpage##1.{\global\pageno@pref##1\global\pageno}}


\hsize=15 cm   \hoffset=0 mm
\vsize=22 cm   \voffset=0 mm

\newdimen\hoffset@corr@p   \newdimen\voffset@corr@p
\newdimen\hoffset@corrm@p   \newdimen\voffset@corrm@p
\newdimen\hoffset@corr@l   \newdimen\voffset@corr@l
\newdimen\hoffset@corrm@l   \newdimen\voffset@corrm@l

\outer\def\portrait{\switch@pl P%
  \glet\hoffset@corr\hoffset@corr@p
  \glet\voffset@corr\voffset@corr@p
  \glet\hoffset@corrm\hoffset@corrm@p
  \glet\voffset@corrm\voffset@corrm@p}
\outer\def\landscape{\switch@pl L%
  \glet\hoffset@corr\hoffset@corr@l
  \glet\voffset@corr\voffset@corr@l
  \glet\hoffset@corrm\hoffset@corrm@l
  \glet\voffset@corrm\voffset@corrm@l}
\def\switch@pl#1{\if #1\ori@pl \else \superendpage \glet\ori@pl#1%
  \dimen@\ht\page@strut \advance\dimen@\dp\page@strut
  \advance\vsize\dimen@ \dimen@ii\hsize \global\hsize\vsize
  \advance\dimen@ii-\dimen@ \global\vsize\dimen@ii \fi}
\let\ori@pl=P

\def\m@g{\dimen@\ht\page@strut \advance\dimen@\dp\page@strut
  \advance\vsize\dimen@ \divide\vsize\count@
  \multiply\vsize\mag \advance\vsize-\dimen@
  \divide\hsize\count@ \multiply\hsize\mag
  \divide\dimen\footins\count@ \multiply\dimen\footins\mag
  \mag\count@}

\output={\physoutput}

\def\physoutput{\make@lbl
  \ifnum \pageno<\firstp@ge \skipp@ge \else
  \ifnum \pageno>\lastp@ge \skipp@ge \else \shipp@ge \fi \fi
  \advancepageno \skippagenum F\skipheadline F\skipfootline F%
  \ifnum\outputpenalty>-\@MM \else \dosupereject \fi}

\def\skippagenum{\glet\skip@page}
\def\skipheadline{\glet\skip@head}
\def\skipfootline{\glet\skip@foot}

\def\skipp@ge{{\skipuserexit \setbox\z@\box\topins
  \setbox\z@\box\footins}\deadcycles\z@}
\def\shipp@ge{\setbox\pageb@x\hbox{%
    \if F\skip@page \pagestyle{\page@pref \folio}\fi}%
  \dimen@-.5\hsize \advance\dimen@\hoffset@corrm
  \divide\dimen@\@m \multiply\dimen@\mag
  \advance\hoffset\dimen@ \advance\hoffset\hoffset@corr
  \dimen@\ht\page@strut \advance\dimen@\dp\page@strut
  \advance\dimen@\vsize \dimen@-.5\dimen@
  \advance\dimen@\voffset@corrm
  \divide\dimen@\@m \multiply\dimen@\mag
  \advance\voffset\dimen@ \advance\voffset\voffset@corr
  \shipout\vbox{\makeheadline \vbadness\@M \setbox\z@\pagebody
    \dimen@\dp\z@ \box\z@ \kern-\dimen@ \makefootline}}

\def\pagecontents{\ifvbox\topins\unvbox\topins\fi
  \dimen@\dp\@cclv \shipuserexit 
  \ifvbox\footins 
    \vskip\skip\footins \footnoterule \unvbox\footins\fi
  \ifr@ggedbottom \kern-\dimen@ \vfil \fi}

\def\folio{\ifnum\pageno<\z@ \ifcase\langu@ge \MEDIUM{\uppercase
  \expandafter{\romannumeral-\pageno}}\else \romannumeral-\pageno \fi
  \else \number\pageno \fi}

\def\makeheadline{\line{\the\headline}\nointerlineskip}
\def\makefootline{\nointerlineskip \line{\the\footline}}

\skippagenum=F   \skipheadline=F   \skipfootline=F

\message{title page macros,}


\outer\def\titlepage{\glet\titl@fill\vfil}
\outer\def\notitlepage{\gdef\titl@fill{\vskip20\p@}}

\newbox\t@pleft   \newbox\t@pright
\def\t@pinit{%
  \global\setbox\t@pleft\vbox{\hrule \@height\z@ \@width.26\hsize}%
  \global\setbox\t@pright\copy\t@pleft}
\t@pinit

\def\topleft{\t@p\t@pleft}
\def\topright{\t@p\t@pright}
\def\t@p#1#2{\global\setbox#1\vtop{\unvbox#1\hbox{\strut #2}}}

\outer\def\submit#1{\topleft{\case@language\submittextone}%
  \topleft{{#1}\case@language\submittexttwo}}

\outer\def\title{\vbox{\line{\box\t@pleft \hss \box\t@pright}}%
  \skippagenum T\skipheadline T\skipfootline T%
  \t@pinit \titl@fill \vskip\chskiptamount \@title}
\let\titcon=\relax  
\outer\def\titcon{\errmessage{please use \noexpand\nl in the title
  instead of \noexpand\titcon}\unchskip \@title}
\def\@title#1{\ctrlines\titlestyle{#1}\chskipl}
\def\titl@#1{\edef\n@xt{\noexpand\@title{#1}}\n@xt}

\def\author{\aut@add\authorstyle}
\def\autcon{\and@con \author}
\def\address{\aut@add\addressstyle}
\def\addcon{\and@con \address}
\def\and@con{\titl@fill \ctrline{\case@language{und\else and}}}

\def\aut@add#1{\titl@fill \ctrlines{#1\use@nl}}
\def\use@nl{\let\\\use@@nl}
\def\use@@nl{\errmessage{please use \noexpand\nl in addresses and
  (lists of) authors instead of \string\\}\nl}

\def\abstract{\titl@fill \he@d{\case@language\abstracthead}%
  \after@arg\titl@fill}

\def\ack{\chskipt \he@d{\case@language\ackhead}}

\def\he@d#1{\ctrline{\headstyle{#1}}\chskipl}

\message{chapters, sections and appendices,}


\newtoks\l@names   \l@names={\\\the@label}
\let\the@label=\empty

\def\label{\num@lett\@label}
\def\@label#1{\def@name\l@names#1{\the@label}}
\def\quote{\num@lett\empty}

\newinsert\lbl@ins
\count\lbl@ins=0   \dimen\lbl@ins=\maxdimen   \skip\lbl@ins=0pt
\newcount\lbln@m   \lbln@m=0
\let\lbl@saved\empty

\def\pagelabel{\num@lett\@pagelabel}
\def\@pagelabel#1{\ifx#1\undefined \let#1\empty \fi
  \toks@\expandafter{#1}\expandafter\testcr@ss\the\toks@\cr@ss\@@
  \ifcr@ss\else
    \toks@{\cr@ss\lbl@undef}\def@name\l@names#1{\the\toks@}\fi
  \g@ne\lbln@m \insert\lbl@ins{\vbox{\vskip\the\lbln@m sp}}%
  \count@\lbln@m \do@label\store@label#1}
\begingroup \let\save=\relax  
  \gdef\lbl@undef{\message{unresolved \string\pagelabel, use
    \string\save\space and \string\crossrestore}??}
\endgroup
\def\do@label#1{\expandafter#1\csname\the\count@\endcsname}
\def\store@label#1#2{\expandafter\gdef\expandafter\lbl@saved
  \expandafter{\lbl@saved#1#2}}

\def\make@lbl{\setbox\z@\vbox{\let\MEDIUM\relax
  \unvbox\lbl@ins \loop \setbox\z@\lastbox \ifvbox\z@
    \count@\ht\z@ \do@label\make@label \repeat}}
\def\make@label#1{\def\make@@label##1#1##2##3#1##4\@@{%
    \gdef\lbl@saved{##1##3}%
    \ifx @##4@\errmessage{This can't happen}\else
    \def@name\l@names##2{\page@pref\folio}\fi}%
  \expandafter\make@@label\lbl@saved#1#1#1\@@}

\outer\def\lblrestore{\all@restore\l@names}


\let\sect=\relax  \let\s@ct=\relax  

\def\chapinit{\chap@init{\chap@pref}\glet\sect@@eq\@chap@sect@eq
  \sectinit}
\def\appinit{\chap@init{\char\the\appn@m}\glet\sect@@eq\@chap@eq
  \glet\sect@dot@pref\empty \glet\sect@pref\dot@pref
  \glet\sect\undefined}
\def\sectinit{\xdef\sect@dot@pref{\the\sectn@m.}%
  \xdef\sect@pref{\dot@pref\sect@dot@pref}\glet\sect\s@ct}
\def\chap@init#1{\xdef\the@label{#1}\xdef\dot@pref{\the@label.}%
  \xdef\dash@pref{\the@label--}\glet\chap@@eq\@chap@eq}


\newcount\chapn@m   \chapn@m=0

\outer\def\chappage{\glet\chap@page T}
\outer\def\nochappage{\glet\chap@page F}

\outer\def\arabicchapnum{\glet\chap@ar A\gdef\chap@pref{\the\chapn@m}}
\outer\def\romanchapnum{\glet\chap@ar R%
  \gdef\chap@pref{\uppercase{\romannumeral\chapn@m}}}

\let\chap=\relax  \let\ch@p=\relax  

\outer\def\chapters{\glet\chap@yn Y\glet\chap\ch@p
  \chap@init{0}\glet\sect@@eq\@chap@sect@eq \sectinit}
\outer\def\nochapters{\glet\chap@yn N\glet\chap\undefined
  \glet\dot@pref\empty \glet\dash@pref\empty
  \glet\chap@@eq\@eq \glet\sect@@eq\@sect@eq \sectinit}

\outer\def\ch@p#1{\if T\chap@page \superendpage \else \chap@break \fi
  \g@ne\chapn@m \sect@reset \chapinit \page@reset\chapn@m
  \eq@reset \fig@reset \tab@reset
  \toks@{\dot@pref}\toks@ii{#1}\chapuserexit
  \titl@{\the\toks@\ \the\toks@ii}%
  \ifnum\auto@toc>\m@ne \toks@store{#1}\@toc\dot@pref \fi}



\def\sec@title#1#2#3#4#5#6#{\ifx @#6@\g@ne#1\else\global#1#6\fi
  #2\secskipt \xdef\the@label{#3\the#1}\xdef#4{\the@label.}%
  \read@store{\sec@@title#4#5}}
\def\sec@@title#1#2#3#4{\toks@{\the@label.}\toks@ii\toks@store
  #4\bpargroup #2\varitem{\the\toks@}\interlinepenalty\@M
    \let\nl\lb \the\toks@ii \par\nobreak
  \ifnum#3<\auto@toc \@toc{#1}\fi}

\newcount\sectn@m   \sectn@m=0
\def\sect@reset{\gz@\sectn@m}
\outer\def\s@ct{\sec@title\sectn@m
  {\secs@reset \sectinit \eq@@reset \fig@@reset \tab@@reset}%
  \dot@pref\sect@pref{\sectstyle\z@\sectuserexit}}

\let\secs@reset=\relax


\def\sect@lev{\@ne}         
\def\sect@id{sect}          
\def\secs@id{secs}          

\outer\def\newsect{\begingroup \count@\sect@lev
  \let\@\endcsname \let\or\relax
  \edef\n@xt{\new@sect}\advance\count@\@ne
  \xdef\sect@lev{\the\count@\space}\n@xt
  \glet\sect@id\secs@id \xdef\secs@id{\secs@id s}\endgroup}
\begingroup \let\newcount=\relax
  \gdef\new@sect{\wlog{\noexpand\string\secs@nm\@= subsection
      level \noexpand\sect@lev}\noexpand\newcount\secs@nm\n@m@
    \gdef\secs@nm\@reset@{\noexpand\gz@\secs@nm\n@m@}%
    \outer\gdef\secs@nm\@{\noexpand\sec@title\secs@nm\n@m@
      \secs@nm s\@reset@ \csn@me\sect@id\@pref@ \secs@nm\@pref@
      {\secs@nm\style@{\the\count@}\secs@nm\userexit@}}%
    \glet\secs@nm s\@reset@ \relax
    \gdef\secs@nm\style@{\csn@me\sect@id\style@}%
    \gdef\secs@nm\userexit@{\secs@nm s\userexit@}%
    \glet\secs@nm s\userexit@\relax
    \outer\xdef\csn@me toc\secs@id\@
      {\global\auto@toc\noexpand\the\count@\space}%
    \xdef\noexpand\save@@toc{\save@@toc\or\secs@id}}
\endgroup

\def\n@m@{n@m\@}
\def\@pref@{@pref\@}
\def\@reset@{@reset\@}
\def\style@{style\@}
\def\userexit@{userexit\@}

\def\secs@nm{\csn@me\secs@id}
\def\csn@me{\expandafter\noexpand\csname}


\newcount\appn@m   \appn@m=64

\outer\def\appendix{\if T\chap@page \superendpage \else\chap@break \fi
  \g@ne\appn@m \secs@reset \appinit \page@reset\appn@m
  \eq@reset \fig@reset \tab@reset \futurelet\n@xt \app@ndix}

\def\app@ndix{\ifcat\bgroup\noexpand\n@xt \expandafter\@ppendix \else
  \expandafter\@ppendix\expandafter\unskip \fi}

\def\@ppendix#1{\toks@{\case@language\appendixhead~\dot@pref}
  \toks@ii{#1}\appuserexit
  \titl@{\the\toks@\ \the\toks@ii}%
  \ifnum\auto@toc>\m@ne \toks@store{#1}%
  \@toc{\case@language\appendixhead\ \dot@pref}\fi}

\def\app#1{\chskipt \he@d{\case@language\appendixhead\ #1}}

\message{equations,}


\def\num@lett{\cat@lett \num@@lett}
\def\num@@lett#1#2{\egroup #1{#2}}

\def\num@l@tt{\cat@lett \num@@l@tt}
\def\num@@l@tt#1#{\egroup #1}

\def\cat@lett{\bgroup
  \catcode`\0\l@tter \catcode`\1\l@tter \catcode`\2\l@tter
  \catcode`\3\l@tter \catcode`\4\l@tter \catcode`\5\l@tter
  \catcode`\6\l@tter \catcode`\7\l@tter \catcode`\8\l@tter
  \catcode`\9\l@tter \catcode`\'\l@tter}

\def\quote@all#1{\leavevmode\hbox{\mathcode`\-\dq 707B$#1$}}
\def\use{\num@lett\@use}
\def\@use{\setbox\z@\hbox}


\newtoks\e@names   \e@names={}
\newcount\eqn@m   \eqn@m=0

\outer\def\equall{\glet\eq@acs A\glet\eq@pref\empty
  \glet\def@eq\@eq \glet\eq@reset\relax \glet\eq@@reset\relax}
\outer\def\equchap{\glet\eq@acs C\gdef\eq@pref{\dot@pref}%
  \gdef\def@eq{\chap@@eq}\glet\eq@reset\eqz@ \glet\eq@@reset\relax}
\outer\def\equsect{\glet\eq@acs S\gdef\eq@pref{\sect@pref}%
  \gdef\def@eq{\sect@@eq}\glet\eq@reset\eqz@ \glet\eq@@reset\eqz@}
\def\eqz@{\gz@\eqn@m}

\outer\def\equfull{\glet\eq@fs F\gdef\eq@@fs{\let\test@eq\full@eq}}
\outer\def\equshort{\glet\eq@fs S\glet\eq@@fs\relax}

\def\@eq(#1){#1}
\def\@chap@eq{\noexpand\chap@eq\@eq}
\def\@sect@eq{\noexpand\sect@eq\@eq}
\def\@chap@sect@eq{\noexpand\chap@sect@eq\@eq}

\def\chap@eq{\test@eq\empty\dot@pref}
\def\sect@eq{\test@eq\empty\sect@dot@pref}
\def\chap@sect@eq{\test@eq\sect@eq\dot@pref}
\def\test@eq#1#2#3.{\def\n@xt{#3.}\ifx#2\n@xt \let\n@xt#1\fi \n@xt}
\def\full@eq#1#2{}
\def\short@eq#1#2#3.{#1}

\outer\def\equleft{\glet\eq@lrn L\glet\eqtag\leqno
  \glet\eq@tag\leq@no}
\outer\def\equright{\glet\eq@lrn R\glet\eqtag\eqno
  \glet\eq@tag\eq@no}
\outer\def\equnone{\glet\eq@lrn N\glet\eqtag\n@eqno
  \glet\eq@tag\neq@no}

\begingroup
  \catcode`\$=\active \catcode`\*=3 \lccode`\*=`\$
  \lowercase{\gdef\n@eqno{\catcode`\$\active
                \def$${\egroup **}\setbox\z@\hbox\bgroup}}
\endgroup
\def\eq@no{\llap{$\@lign##$}\tabskip\z@skip}
\def\leq@no{\kern-\displaywidth \rlap{$\@lign##$}\tabskip\displaywidth}
\def\neq@no{\@use{$\@lign##$}\tabskip\z@skip}

\def\displaylines{\afterassignment\display@lines \@eat}
\def\display@lines{\displ@y
  \halign\n@xt\hbox to\displaywidth{$\@lign\hfil\displaystyle##\hfil$}%
    &\span\eq@tag\crcr}

\def\eqalignno{\let\eq@@tag\eq@no \eqalign@tag}
\def\leqalignno{\let\eq@@tag\leq@no \eqalign@tag}
\def\eqaligntag{\let\eq@@tag\eq@tag \eqalign@tag}
\def\eqalign@tag{\afterassignment\eqalign@@tag \@eat}
\def\eqalign@@tag{\displ@y
  \tabskip\centering \halign to\displaywidth\n@xt
    \hfil$\@lign\displaystyle{##}$\tabskip\z@skip
    &$\@lign\displaystyle{{}##}$\hfil\tabskip\centering
    &\span\eq@@tag\crcr}

\def\fulltag#1{{\let\test@eq\full@eq#1}}
\def\shorttag#1{{\let\test@eq\short@eq#1}}

\def\eq{\g@ne\eqn@m \make@eq\empty}
\def\make@eq#1{(\eq@pref\the\eqn@m #1)}
\def\EQ{\eq \num@lett\eq@save}
\def\eq@save#1{\def@name\e@names#1{\expandafter\def@eq\make@eq\empty}}

\def\eqn{\eqtag\eq}
\def\EQN{\eqtag\EQ}

\def\eqadv{\g@ne\eqn@m}
\def\EQADV{\eqadv \num@lett\eq@save}

\newcount\seqn@m   \seqn@m=96

\def\subeqbegin{\global\seqn@m96 \subeq}
\def\SUBEQBEGIN{\global\seqn@m96 \SUBEQ}
\def\subeq{\g@ne\seqn@m \make@eq{\char\seqn@m}}
\def\SUBEQ{\num@lett\@SUBEQ}
\def\@SUBEQ#1{\subeq \def@name\e@names#1{\char\the\seqn@m}}

\def\SUBEQNBEGIN{\eqtag\SUBEQBEGIN}

\def\SUBEQN{\eqtag\SUBEQ}

\def\eqapp{\num@lett\@eqapp}
\def\@eqapp#1#2{(\fulltag#1#2)}

\def\queq{\num@lett\@queq}
\def\@queq#1{\quote@all{\eq@@fs(#1)}}
\def\qeq{\case@abbr\eqabbr\queq}
\def\qeqs{\case@abbr\eqsabbr\queq}

\outer\def\eqrestore{\all@restore\e@names}

\message{storage management,}



\newtoks\toks@store
\newtoks\file@list \file@list={1234567}
\def\@tmp{\jobname.$$}
\let\ext@ft=F

\begingroup
  \let\storebox=\relax \let\refnam=\relax  
  \let\RFfile=\relax \let\RFext=\relax     
  \newhelp\opt@help{The options \string\refnam, \string\RFfile\space
    and \string\RFext\space are incompatible with \string\storebox.
    Your request will be ignored.}
  \global\opt@help=\opt@help 
  \gdef\opt@err{{\errhelp\opt@help \errmessage{Incompatible options}}}
\endgroup

\begingroup
  \let\storebox=\relax \let\storelist=\relax  
  \let\storefile=\relax \let\RFfile=\relax    
  \gdef\case@store{%
    \glet\storebox\undefined
    \if B\store@blf \glet\storebox\empty \glet\case@store\case@box
      \else \glet\box@store\undefined
      \glet\box@out\undefined \glet\box@print\undefined
      \glet\box@save\undefined \glet\box@kill\undefined \fi
    \glet\storelist\undefined
    \if L\store@blf \glet\storelist\empty \glet\case@store\case@list
      \else \glet\list@store\undefined
      \glet\list@out\undefined \glet\list@print\undefined
      \glet\list@save\undefined \glet\list@kill\undefined \fi
    \glet\storefile\undefined
    \if F\store@blf \glet\storefile\empty \glet\case@store\case@file
      \else \store@setup
      \glet\file@out\undefined \glet\file@print\undefined
      \glet\filef@rm@t\undefined \glet\fil@f@rm@t\undefined
      \glet\file@save\undefined \glet\file@kill\undefined \fi
    \glet\case@box\undefined \glet\case@list\undefined
    \glet\case@file\undefined \glet\store@setup\undefined
    \case@store}
  \gdef\store@setup{\ifx \RFfile\undefined \glet\file@store\undefined
    \glet\file@open\undefined \glet\file@close\undefined
    \glet\file@wlog\undefined \glet\file@free\undefined
    \glet\file@copy\undefined \glet\file@read\undefined \fi}
\endgroup

\outer\def\storebox{\if T\ext@ft \opt@err \else
    \if L\RF@lfe \if N\ref@sbn \opt@err
    \else \glet\store@blf B\fi \else \opt@err \fi \fi}
\outer\def\storelist{\glet\store@blf L}
\outer\def\storefile{\glet\store@blf F}

\def\read@store{\bgroup \@read@store}
\def\read@@store{\bgroup \catcode`\@\l@tter \@read@store}
\def\@read@store#1{\def\after@read{\egroup \toks@store\toks@i
    #1\after@read \ignorespaces}%
  \catcode`\^^M\active \afterassignment\after@read \global\toks@i}
\def\afterread#1{\bgroup \def\after@read{\egroup #1\after@read}}
\let\after@read=\relax
\begingroup
  \catcode`\:=\active
  \gdef\write@save#1{\write@store{:restore\@type{#1}}}
\endgroup
\def\write@store#1{\s@ve{#1{\the\toks@store}}}

\def\f@rm@t#1#2{\bgroup \ignorefoot
  \leftskip\z@skip \rightskip\z@skip \f@rmat
  \ifx @#1@\everypar{\b@format}\else
    \varitem\@indent{#1}\b@format \fi #2\e@format}
\let\f@rmat=\nointerlineskip
\def\form@t{\unskip \strut \par \@break \eform@t}
\def\b@format{\glet\e@format\form@t \strut}
\def\eform@t{\egroup \glet\e@format\eform@t}
\let\e@format=\eform@t

\def\@store{\case@store\box@store\list@store\file@store}
\def\@sstore#1#2#3{\par \noindent \bgroup \captionstyle
  \case@abbr#2#3:\enskip \the\toks@store \par \egroup \@store#1{#3.}}
\def\@add#1{\read@store{\@store#1{}}}
\def\@out#1#2#3#4#5{\case@store\box@out\list@out\file@out#1\begingroup
  \if T#2\let\chap@break\superendpage \fi \chap@break
  \chap@init{\case@language#3}%
  \if C\page@ac \skippagenum T\fi
  \page@reset6#1%
  \@style \he@d{\strut\case@language#4}\@break \@print#1%
  \ifx\chap@break\superendpage \superendpage \fi
  \def\\##1{\glet##1\undefined}\the#5\global#5\emptyt@ks
  \endgroup \fi}
\def\@print{\case@store\box@print\list@print\file@print}
\def\@save{\case@store\box@save\list@save\file@save}
\def\@kill{\case@store\box@kill\list@kill\file@kill}
\def\@ext#1#2#3 {\if B\store@blf \opt@err \else
  \glet\ext@ft T\@add#1{#2#3 }\bgroup
   \def\@store##1##2{}\input#3 \egroup \fi}
\def\@@ext#1#2{\let#1#2\everypar\emptyt@ks
  \def\read@store##1{\relax##1}\def\@store##1{\f@rm@t}\input}

\def\case@box#1#2#3{\case@@store#1%
  \fig@box\tab@box\ref@box\toc@box\foot@box}
\def\case@list#1#2#3{\case@@store#2%
  \fig@list\tab@list\ref@list\toc@list\foot@list}
\def\case@file#1#2#3{\case@@store{\expandafter#3}%
  \fig@file\tab@file\ref@file\toc@file\foot@file}

\def\case@@store#1#2#3#4#5#6#7{\ifcase#7%
  \toks@{\fig@type#1#2}\or
  \toks@{\tab@type#1#3}\or
  \toks@{\ref@type#1#4}\or
  \toks@{\toc@type#1#5}\or
  \toks@{\foot@type#1#6}\fi
  \expandafter\let\expandafter\@type\the\toks@}
\def\@style{\csname\@type style\endcsname}
\def\@indent{\csname\@type indent\endcsname}
\def\@break{\csname\@type break\endcsname}

\def\box@store#1#2{\global\setbox#1\vbox
  {\ifvbox#1\unvbox#1\fi \@style \f@rm@t{#2}{\the\toks@store}}}
\def\box@out{\ifvbox}
\def\box@print{\vskip\baselineskip \unvbox}
\begingroup
  \catcode`\:=\active \catcode`\;=\active
  \gdef\box@save#1{\wlog{; Unable to save text for
    \@type's with option :storebox}}
\endgroup
\def\box@kill#1{{\setbox\z@\box#1}}

\def\list@store#1#2{\toks@\expandafter{#1\\}%
  \xdef#1{\the\toks@ {#2}{\the\toks@store}}}
\def\list@out#1{\ifx #1\empty \else}
\def\list@print#1{\let\\\f@rm@t #1\glet#1\empty}
\def\list@save{\def\\##1##2{\toks@store{##2}\write@save{##1}}%
  \newlinechar`\^^M}
\def\list@kill#1{\glet#1\empty}

\begingroup
  \catcode`\:=\active
  \gdef\file@store#1#2#3#4{%
    \if0#3\expandafter\file@open\the\file@list\@@#1#2\fi
    {\newlinechar`\^^M\let\save@write#2\write@store{::{#4}}}}
\endgroup
\def\file@open#1#2\@@#3#4{\immediate\openout#4\@tmp#1
  \gdef#3{#3#4#1}\global\file@list{#2}\file@wlog{open}#1}
\def\file@wlog#1#2{\wlog{#1 \@tmp#2 for \@type's}}
\def\file@out#1#2#3{\if0#3\else}
\def\file@print#1#2#3{\file@close#2#3\let\\\filef@rm@t
  \file@copy#1#2#3}
\def\filef@rm@t#1{\bgroup \catcode`\@\l@tter \fil@f@rm@t{#1}}
\def\fil@f@rm@t#1#2{\egroup \f@rm@t{#1}{#2}}
\def\file@save#1#2#3{\if0#3\else \file@close#2#3%
  \def\\##1{\read@@store{\expandafter\file@store#1{##1}%
    \write@save{##1}}}%
  \newlinechar`\^^M\file@copy#1#2#3\fi}
\def\file@kill#1#2#3{\if0#3\else \file@close#2#3\file@free#1#2#3\fi}
\def\file@close#1#2{\immediate\closeout#1\file@wlog{close}#2}
\def\file@copy#1#2#3{\file@free#1#2#3\file@read#3}
\def\file@read#1{\input\@tmp#1 }
\def\file@free#1#2#3{\gdef#1{#1#20}%
  \global\file@list\expandafter{\the\file@list#3}}

\message{figures,}


\def\if@t#1#2#3#4#5#6{\glet#1#6\gdef#2{\dot@pref}\glet#3#5\glet#4\relax
  \if#6A\glet#2\empty \glet#3\relax \fi
  \if#6S\gdef#2{\sect@pref}\glet#4#5\fi}
\def\bf@t#1{\let\f@t#1\num@l@tt}
\def\ef@t#1#2#3#4#{\ifx @#4@\g@ne#1\xdef\thef@tn@m{#2\the#1}\else
  \gdef\thef@tn@m{#4}\fi #3\read@store\f@t}
\def\@extf@t#1#2#{\gdef\thef@tn@m{#1}\f@t}


\newtoks\f@names   \f@names={}
\newcount\fign@m   \fign@m=0
\def\fig@type{fig}
\newbox\fig@box
\let\fig@list=\empty
\newwrite\fig@write  \def\fig@file{\fig@file\fig@write0}

\outer\def\figall{\fig@init A}
\outer\def\figchap{\fig@init C}
\outer\def\figsect{\fig@init S}
\def\fig@init{\if@t\fig@acs\fig@pref\fig@reset\fig@@reset\figz@}
\def\figz@{\gz@\fign@m}

\outer\def\figpage{\glet\fig@page T}
\outer\def\nofigpage{\glet\fig@page F}

\def\fig{\bf@t\fig@\@fig}
\def\FIG{\bf@t\fig@\@FIG}
\def\ffig{\bf@t\ffig@\@fig}
\def\FFIG{\bf@t\ffig@\@FIG}

\def\@fig{\ef@t\fign@m\fig@pref\relax}
\def\@FIG#1{\ef@t\fign@m\fig@pref{\def@name\f@names#1{\thef@tn@m}}}

\def\fig@{\@store0{\thef@tn@m .}}
\def\ffig@{\@sstore0\figpref{\thef@tn@m}}
\def\figadd{\@add0}

\def\qufig{\case@abbr\figabbr\num@lett\quote@all}
\def\qufigs{\case@abbr\figsabbr\num@lett\quote@all}

\outer\def\figout{\@out0\fig@page\figpref\fighead\emptyt@ks}
\outer\def\figkill{\@kill0}
\outer\def\restorefig#1{\read@store{\@store0{#1}}}
\outer\def\figrestore{\all@restore\f@names}

\outer\def\FIGext{\@ext0\FIG@ext}
\def\FIG@ext{\@@ext\@FIG\@extf@t}


\newdimen\spictskip  \spictskip=2.5pt

\def\pict{\bf@t\pict@\@fig}
\def\PICT{\bf@t\pict@\@FIG}

\def\pict@{\vskip\the\toks@store \bpargroup
  \raggedright \captionstyle \varitem{\qufig{\thef@tn@m\,}:}%
  \let\@spict\spict@ \spacefactor998\ignorespaces}

\def\spict#1{\ifnum\spacefactor=998\else \parvskip\spictskip \fi
  \@spict{#1\enskip}\ignorespaces}
\def\spict@#1{\setbox\z@\hbox{#1}\advance\hangindent\wd\z@
  \box\z@ \let\@spict\llap}


\outer\def\graphics{\glet\graphic\gr@phic}
\outer\def\nographics{\glet\graphic\nogr@phic}

\def\gr@phic#1{\vbox\bgroup \def\gr@@@ph{#1}\bfr@me\gr@ph}
\def\nogr@phic#1{\vbox\bgroup \write\m@ne{Insert plot #1}\bfr@me\fr@me}
\def\frame{\vbox\bgroup \bfr@me\fr@me}
\def\bfr@me#1#2#3#4{\tfr@me\z@\dimen@iv#2\relax 
  \dimen@#3\relax \tfr@me\dimen@\dimen@ii#4\relax 
  \setbox\z@\hbox to\dimen@iv{#1}\ht\z@\dimen@ \dp\z@\dimen@ii
  \box\z@ \egroup}
\def\tfr@me#1#2#3\relax{#2#3\relax \ifdim#2<-#1\errhelp\fr@mehelp
  \errmessage{Invalid box size}#2-#1\fi}
\newhelp\fr@mehelp{The \string\wd\space and \string\ht+\string\dp\space
  of a \string\frame\space or \string\graphic\space \string\box\space
  must not be negative and will be changed to 0pt.}

\def\gr@ph{\lower\dimen@ii\gr@@ph b\hfil
  \raise\dimen@\gr@@ph e}
\def\gr@@ph#1{\hbox{\special{^X\gr@@@ph^A}}}         
\def\gr@@ph#1{\hbox{\special{#1plot GKSM \gr@@@ph}}} 
\def\gr@@ph#1{\hbox{\special{^X#1plot GKSM \gr@@@ph^A}}} 
\def\fr@me{\vrule\fr@@@me
  \bgroup \dimen@ii-\dimen@ \fr@@me\dimen@ii \hfilneg
  \bgroup \dimen@-\dimen@ii \fr@@me\dimen@ \vrule\fr@@@me}
\def\fr@@me#1{\advance#1.4\p@ \leaders \hrule\fr@@@me \hss \egroup}
\def\fr@@@me{\@height\dimen@ \@depth\dimen@ii}

\message{tables,}


\newtoks\t@names   \t@names={}
\newcount\tabn@m   \tabn@m=0
\def\tab@type{tab}
\newbox\tab@box
\let\tab@list=\empty
\newwrite\tab@write  \def\tab@file{\tab@file\tab@write0}

\outer\def\taball{\tab@init A}
\outer\def\tabchap{\tab@init C}
\outer\def\tabsect{\tab@init S}
\def\tab@init{\if@t\tab@acs\tab@pref\tab@reset\tab@@reset\tabz@}
\def\tabz@{\gz@\tabn@m}

\outer\def\tabpage{\glet\tab@page T}
\outer\def\notabpage{\glet\tab@page F}

\def\tab{\bf@t\tab@\@tab}
\def\TAB{\bf@t\tab@\@TAB}
\def\ttab{\bf@t\ttab@\@tab}
\def\TTAB{\bf@t\ttab@\@TAB}

\def\@tab{\ef@t\tabn@m\tab@pref\relax}
\def\@TAB#1{\ef@t\tabn@m\tab@pref{\def@name\t@names#1{\thef@tn@m}}}

\def\tab@{\@store1{\thef@tn@m .}}
\def\ttab@{\@sstore1\tabpref{\thef@tn@m}}
\def\tabadd{\@add1}

\def\qutab{\case@abbr\tababbr\num@lett\quote@all}
\def\qutabs{\case@abbr\tabsabbr\num@lett\quote@all}

\outer\def\tabout{\@out1\tab@page\tabpref\tabhead\emptyt@ks}
\outer\def\tabkill{\@kill1}
\outer\def\restoretab#1{\read@store{\@store1{#1}}}
\outer\def\tabrestore{\all@restore\t@names}

\outer\def\TABext{\@ext1\TAB@ext}
\def\TAB@ext{\@@ext\@TAB\@extf@t}


\newskip\htabskip   \htabskip=1em plus 2em minus .5em
\newdimen\vtabskip  \vtabskip=2.5pt
\newbox\tab@top   \newbox\tab@bot

\let\@hrule=\hrule
\let\@halign=\halign
\let\@valign=\valign
\let\@span=\span
\let\@omit=\omit

\def\@@span{\@span\@omit\@span}
\def\@@@span{\@span\@omit\@@span}

\def\sp@n{\span\@omit\advance\mscount\m@ne} 

\def\table#1#{\vbox\bgroup\offinterlineskip
  \toks@ii{#1\bgroup \unhcopy\tab@top \unhcopy\tab@bot
    ##}\afterassignment\tab@preamble \@eat}


\def\tab@preamble#1\cr{\let\tab@@vrule\tab@repeat
  \let\tab@amp@\empty \let\tab@amp\empty \let\span@\@@span
  \toks@{\tab@space#1&\cr}\the\toks@}
\def\tab@space{\tab@test{ }{}\tab@vrule}                 
\def\tab@vrule{\tab@test\vrule{\tab@add\vrule}
  \tab@@vrule}
\def\tab@repeat{\tab@test&{\tab@add&
    \let\tab@@vrule\tab@template \let\tab@amp@\tab@@amp
    \let\tab@amp&\let\span@\@@@span}\tab@template}
\def\tab@template#1&{\tab@add{\@span\tab@amp@
    \tabskip\htabskip&\tab@setup#1&\tabskip\z@skip##}
  \tab@test\cr{\let\tab@@vrule\tab@exec}\tab@space}      
\def\tab@@amp{&##}

\def\tab@test#1#2#3{\let\tab@comp= #1\toks@{#2}\let\tab@go#3%
  \futurelet\n@xt \tab@@test}
\def\tab@@test{\ifx \tab@comp\n@xt \the\toks@
    \afterassignment\tab@go \expandafter\@eat \else
  \expandafter\tab@go \fi}

\def\tab@add#1{\toks@ii\expandafter{\the\toks@ii#1}}


\def\tab@exec{\tab@r@set\everycr{\tab@body}%
  \def\halign{\tab@r@set \halign}\def\valign{\tab@r@set \valign}%
  \def\omit{\@omit \tab@setup}%
  \def\n@xt##1{\hbox{\dimen@\ht\strutbox\dimen@ii\dp\strutbox
    \advance##1\vtabskip \vrule \@height\dimen@ \@depth\dimen@ii
    \@width\z@}}%
  \setbox\tab@top\n@xt\dimen@ \setbox\tab@bot\n@xt\dimen@ii
  \def\ml##1{\relax                                      
    \ifmmode \let\@ml\empty \else \let\@ml$\fi           
    \@ml\vcenter{\hbox\bgroup\unhcopy\tab@top            
    ##1\unhcopy\tab@bot\egroup}\@ml}%
  \def\nl{\egroup\hbox\bgroup\strut}
  \tabskip\z@skip\@halign\the\toks@ii\cr}

\def\tab@r@set{\let\cr\endline \everycr\emptyt@ks
  \let\halign\@halign \let\valign\@valign
  \let\span\@span \let\omit\@omit}
\def\tab@setup{\relax \iffalse {\fi \let\span\span@ \iffalse }\fi}


\def\tab@body{\noalign\bgroup \tab@@body}           
\def\tab@@body{\futurelet\n@xt \tab@end}            
\def\tab@end{\ifcat\egroup\noexpand\n@xt
    \expandafter\egroup \expandafter\egroup \else        
  \expandafter\tab@blank \fi}
\def\tab@blank{\ifcat\space\noexpand\n@xt
    \afterassignment\tab@@body \expandafter\@eat \else   
  \expandafter\tab@hrule \fi}
\def\tab@hrule{\ifx\hrule\n@xt
    \def\hrule{\@hrule\egroup \tab@body}\else            
  \expandafter\tab@noalign \fi}
\def\tab@noalign{\ifx\noalign\n@xt
    \aftergroup\tab@body \expandafter\@eat@ \else        
  \expandafter\tab@row \fi}
\def\tab@row#1\cr{\toks@ii{\toks@ii{\egroup}
    \tab@item#1&\cr}\the\toks@ii}
\def\tab@item#1&{\tab@add{\tab@amp&#1&}
  \futurelet\n@xt \tab@cr}
\def\tab@cr{\ifx\cr\n@xt
    \expandafter\the\expandafter\toks@ii \else           
  \expandafter\tab@item \fi}

\message{references,}


\newtoks\r@names   \r@names={}
\newcount\refn@m   \refn@m=0
\newcount\ref@temp
\def\ref@type{ref}
\newbox\ref@box
\let\ref@list=\empty
\newwrite\ref@write  \def\ref@file{\ref@file\ref@write0}

\begingroup \let\refnam=\relax  
  \newhelp\ref@help{The option \string\refnam\space allows predefined
    references only and is incompatible with \string\qref(s).
    Your request will be ignored.}
  \global\ref@help=\ref@help 
  \gdef\ref@err{{\errhelp\ref@help \errmessage{Invalid request}}}
\endgroup

\begingroup
  \let\refsup=\relax \let\refsqb=\relax  
  \let\refnam=\relax                     
  \gdef\ref@setup{%
    \glet\refsup\undefined
    \if S\ref@sbn \glet\refsup\empty \glet\therefn@m\suprefn@m \fi
    \glet\refsqb\undefined
    \if B\ref@sbn \glet\refsqb\empty \glet\therefn@m\sqbrefn@m \fi
    \glet\refnam\undefined
    \if N\ref@sbn \glet\refnam\empty \glet\the@quref\nam@quref
      \glet\@@ref\ref@err \gdef\qref{\ref@err \quref}\glet\qrefs\qref
      \glet\RF@def@\RF@def@nam
      \glet\RF@find\undefined \glet\RF@search\undefined
      \glet\RF@locate\undefined \glet\@RFread\undefined
      \glet\qurefsup\ref@err \glet\sup@quref\undefined
      \glet\qurefsqb\ref@err \glet\sqb@quref\undefined
      \glet\qurefnum\ref@err \glet\num@quref\undefined
      \glet\ref@restore\ref@err
      \else \gdef\@@ref{\@store2\therefn@m}%
      \gdef\qref{\case@abbr\refabbr\num@lett\quote@all}%
      \gdef\qrefs{\case@abbr\refsabbr\num@lett\quote@all}%
      \glet\RF@def@\RF@def@num \glet\RF@print\undefined
      \gdef\ref@restore{\all@restore\r@names}\fi
    \glet\nam@quref\undefined
    \gdef\quref{\ref@unskip \num@lett\the@quref}%
    \glet\RF@def@num\undefined \glet\RF@def@nam\undefined
    \gdef\RF@restore{\all@restore\R@names}%
    \glet\ref@setup\undefined}
\endgroup

\outer\def\refsup{\glet\ref@sbn S\global\qurefsup}
\outer\def\refsqb{\glet\ref@sbn B\global\qurefsqb}
\outer\def\refnam{\if B\store@blf \opt@err \else \glet\ref@sbn N\fi}

\def\qurefsup{\let\the@quref\sup@quref}
\def\qurefsqb{\let\the@quref\sqb@quref}
\def\qurefnum{\let\the@quref\num@quref}

\outer\def\refpage{\glet\ref@page T}
\outer\def\norefpage{\glet\ref@page F}

\outer\def\refkeep{\glet\ref@kc K}
\outer\def\refclear{\glet\ref@kc C}

\def\ref{\ref@advance \refend \@ref}
\def\REF{\num@lett\@REF}
\def\@REF#1{\ref@name#1\@ref}
\def\refend{\quref{\the\refn@m}}

\def\refs{\ref@advance \ref@temp\refn@m \@ref}
\def\REFS{\num@lett\@REFS}
\def\@REFS#1{\ref@name#1\ref@temp\refn@m \@ref}
\def\refscon{\ref@advance \@ref}

\def\refsend{\quref{\the\ref@temp -\the\refn@m}}

\def\ref@advance{\ref@unskip \g@ne\refn@m}
\def\ref@name{\ref@@name\r@names}
\def\ref@@name#1#2{\ref@advance \def@name#1#2{\the\refn@m}}
\def\ref@unskip{\ifhmode \unskip \fi}

\def\suprefn@m{\the\refn@m .}
\def\sqbrefn@m{$\lbrack \the\refn@m \rbrack$}
\def\@ref{\read@store\@@ref}
\def\@@ref{\ref@setup \@@ref}
\def\refadd{\@add2}

\def\sup@quref#1{\leavevmode \nobreak \quote@all{^{#1}}}
\def\sqb@quref#1{\ \quote@all{\lbrack #1\rbrack}}
\def\num@quref{\ \quote@all}
\def\nam@quref{\@use}
\def\quref{\ref@setup \quref}
\def\qref{\ref@setup \qref}
\def\qrefs{\ref@setup \qrefs}

\outer\def\refout{{\if K\ref@kc \@out2\ref@page\refpref\refhead\emptyt@ks
  \else \@out2\ref@page\refpref\refhead\r@names
  \let\\\@RFdef \the\R@names \global\R@names\emptyt@ks
  \if L\RF@lfe \else \glet\RF@list\empty \gz@\RF@high \fi
  \gz@\refn@m \fi}}
\outer\def\refkill{\@kill2}
\outer\def\restoreref#1{\read@@store{\@store2{#1}}}
\outer\def\refrestore{\ref@restore}
\outer\def\RFrestore{\RF@restore}
\def\ref@restore{\ref@setup \ref@restore}
\def\RF@restore{\ref@setup \RF@restore}

\outer\def\REFext{\@ext2\REF@ext}
\def\REF@ext{\@@ext\ref@name\refn@m}


\newtoks\R@names   \R@names={}
\newcount\RFn@m   \newcount\RF@high
\newcount\RFmax   \RFmax=50  
\def\RF@type{RF}
\let\RF@list=\empty
\newwrite\RF@write  \def\RF@file{\RF@file\RF@write0}
\let\RF@noc=N

\begingroup \let\storefile=\relax        
  \let\RFlist=\relax \let\RFfile=\relax  
  \let\RFext=\relax \let\RF=\relax       
  \gdef\@RF{%
    \glet\RFlist\undefined
    \if L\RF@lfe \glet\RFlist\empty \glet\@RF\@RFlist
      \gdef\RF@input{\RF@list}%
      \else \glet\@RFlist\undefined \fi
    \glet\RFfile\undefined
    \if F\RF@lfe \glet\RFfile\empty \glet\@RF\@RFfile
      \else \RF@setup
      \glet\@RFfile\undefined \glet\@RFcopy\undefined
      \glet\RF@store\undefined \glet\RF@copy\undefined
      \glet\RF@@input\undefined \fi
    \glet\RFext\undefined
    \if E\RF@lfe \gdef\RFext##1 {}\glet\@RF\@RFext
      \else \glet\@RFext\undefined \fi
    \glet\RF@setup\undefined \@RF}
  \gdef\RF@setup{\ifx \storefile\undefined \glet\file@store\undefined
    \glet\file@open\undefined \glet\file@close\undefined
    \glet\file@wlog\undefined \glet\file@free\undefined
    \glet\file@copy\undefined \glet\file@read\undefined \fi}
\endgroup

\outer\def\RFlist{\glet\RF@lfe L}
\outer\def\RFfile{\if B\store@blf \opt@err \else \glet\RF@lfe F\fi}
\outer\def\RFext#1 {\if B\store@blf \opt@err \else
  \glet\RF@lfe E\gdef\RF@input{\input#1 }\RF@input \fi}

\def\@RFdef#1{\gdef#1{\RF@def#1}}
\def\@RF@list#1{\toks@\expandafter{\RF@list\RF@#1}%
  \xdef\RF@list{\the\toks@ {\the\toks@store}}}
\def\@RFcopy#1{\RF@store{\noexpand#1}}
\def\RF@store{\let\@type\RF@type \if C\RF@noc \glet\RF@noc N%
  \RF@copy \fi \glet\RF@noc O\expandafter\file@store\RF@file}
\def\RF@input{\expandafter\RF@@input\RF@file}
\begingroup \let\RF=\relax  
  \gdef\RF@copy{{\let\\\RF \let\@RF\@RFcopy
    \expandafter\file@copy\RF@file}}
  \gdef\RF@@input#1#2#3{\if O\RF@noc \let\@type\RF@type
    \file@close#2#3\glet\RF@noc C\fi \let\\\RF \file@read#3}
\endgroup

\def\RF@def#1{\ref@@name\R@names#1\RF@def@ #1}
\def\RF@def@{\ref@setup \RF@def@}
\def\RF@def@num{\toks@store\expandafter{\expandafter\RF@find
  \expandafter{\the\refn@m}}\@@ref}
\def\RF@def@nam{\ifnum\refn@m=\@ne \refadd{\RF@print}\fi}
\def\RF@test#1{\z@}

\def\RF@print{\let\@RF\@RF@print \let\RF@def\RF@test
  \let\RF@first T\RF@input}
\def\@RF@print#1{\ifnum#1>\z@
  \if\RF@first T\let\RF@first F\setbox\z@\lastbox
  \else \form@t\f@rmat \noindent \strut \fi
  \hangindent\namrefindent \the\toks@store \fi}

\def\RF@find#1{\RFn@m#1\bgroup \let\RF@def\RF@test
  \if L\RF@lfe \else \ifnum\RFn@m<\RF@high \else \RF@search \fi \fi
  \let\RF@\RF@locate \RF@list \egroup}

\def\RF@search{\global\RF@high\RFn@m \global\advance\RF@high\RFmax
  \glet\RF@list\empty \let\@RF\@RFread \let\par\relax \RF@input}
\def\@RFread#1{\ifnum#1<\RFn@m \else \ifnum#1<\RF@high
  \@RF@list#1\fi \fi}
\def\RF@locate#1#2{\ifnum#1=\RFn@m #2\fi}

\def\@RFlist#1{\@RFdef#1\@RF@list#1}
\def\@RFfile#1{\@RFdef#1\@RFcopy#1}
\let\@RFext=\@RFdef

\outer\def\RF{\num@lett\RF@}
\def\RF@#1{\ref@unskip \read@store{\@RF#1}}


\outer\def\yearpage{\glet\yearpage@yp Y}
\outer\def\pageyear{\glet\yearpage@yp P}

\def\journal#1{{\journalstyle{#1}}\j@urnal{}}
\def\journalp#1{{\journalstyle{#1}}\j@urnal}
\def\journalf#1#2#3({{\journalstyle{#1}}\j@urnal{#3}#2(}

\def\j@urnal#1#2(#3)#4*{\unskip
  \ {\volumestyle{#1\ifx @#1@\else\ifx @#2@\else
  \kern.2em\fi \fi#2}}\unskip
  \ifx @#3@\else\ifx @#4@ (#3)\else\if Y\yearpage@yp\ (#3) #4\else
  , #4 (#3)\fi \fi \fi}


\message{table of contents,}


\newcount\tocn@m   \tocn@m=0
\newcount\auto@toc   \auto@toc=-1
\let\toc@saved\empty

\def\toc@type{toc}
\newbox\toc@box
\let\toc@list=\empty
\newwrite\toc@write  \def\toc@file{\toc@file\toc@write0}

\outer\def\tocpage{\glet\toc@page T}
\outer\def\notocpage{\glet\toc@page F}

\outer\def\tocnone{\gm@ne\auto@toc}
\outer\def\tocchap{\gz@\auto@toc}
\outer\def\tocsect{\global\auto@toc\@ne}

\def\toc#1{\read@store{\@toc{#1}}}
\def\tocadd{\@add3}
\def\@toc{\g@ne\tocn@m
  \expandafter\@@toc\csname toc@\romannumeral\tocn@m\endcsname}
\def\@@toc#1{\pagelabel#1%
  \toks@store\expandafter{\the\toks@store\toc@fill#1}\@store3}
\def\toc@fill{\rightskip4em\@plus1em\@minus1em\parfillskip-\rightskip
  \unskip\vadjust{}\leaders\hbox to1em{\hss.\hss}\hfil}

\outer\def\tocout{\@out3\toc@page\tocpref\tochead\emptyt@ks}
\outer\def\tockill{\@kill3}
\outer\def\restoretoc#1{\read@@store{\@store3{#1}}}

\message{footnotes,}


\newcount\footn@m   \footn@m=0
\def\foot@type{foot}
\newbox\foot@box
\let\foot@list=\empty
\newwrite\foot@write  \def\foot@file{\foot@file\foot@write0}

\outer\def\footsqb{\glet\foot@bp B\glet\thefootn@m\sqbfootn@m}
\outer\def\footpar{\glet\foot@bp P\glet\thefootn@m\parfootn@m}

\outer\def\footbot{\glet\foot@be B\glet\vfootnote\vfootn@te}
\outer\def\footend{\glet\foot@be E\glet\vfootnote\foot@store}

\outer\def\footpage{\glet\foot@page T}
\outer\def\nofootpage{\glet\foot@page F}

\def\sqbfootn@m{\lbrack \the\footn@m \rbrack}
\def\parfootn@m{\the\footn@m )}

\def\foot{\hfoot \vfootnote\footid}
\def\hfoot{\g@ne\footn@m \edef\n@xt{{$^{\thefootn@m}$}}%
  \expandafter\hfootnote\n@xt}
\def\footnote#1{\hfootnote{#1}\vfootnote\footid}
\def\hfootnote#1{\let\@sf\empty
  \ifhmode\unskip\edef\@sf{\spacefactor\the\spacefactor}\/\fi
  #1\@sf \gdef\footid{#1}}
\def\vfootn@te#1{\insert\footins\bgroup \foot@style
  \llap{#1}\after@arg\@foot}
\let\fo@t=\undefined
\let\f@@t=\undefined
\let\f@t=\undefined

\def\foot@style{\footstyle  
  \interlinepenalty\interfootnotelinepenalty
  \baselineskip\footnotebaselineskip
  \splittopskip\interfootnoteskip 
  \splitmaxdepth\dp\strutbox \floatingpenalty\@MM
  \leftskip.05\hsize \rightskip\z@skip
  \spaceskip\z@skip \xspaceskip\z@skip \noindent \footstrut}

\def\foot@store#1{\read@store{\@store4{#1}}}
\def\footadd{\@add4}

\outer\def\footout{\@out4\foot@page\footpref\foothead\emptyt@ks}
\outer\def\footkill{\@kill4}
\outer\def\restorefoot#1{\read@store{\@store4{#1}}}

\def\ignorefoot{\let\foot\eat \let\hfoot\eat  
  \def\footnote{\expandafter\eat\eat}\let\hfootnote\footnote}

\message{items and points,}



\def\varitem{\afterassignment\v@ritem \setbox\z@\hbox}
\def\v@ritem{\hss \bgroup \aftergroup\v@@ritem}
\def\v@@ritem{\enskip \egroup \endgraf \noindent
  \hangindent\wd\z@ \box\z@ \ignorespaces}

\def\hvskip{\afterassignment\h@vskip \skip@}
\def\h@vskip{\unskip\nobreak \vadjust{\vskip\skip@}\lb \ignorespaces}

\def\parvskip{\bgroup \afterassignment\par@vskip \parskip}
\def\par@vskip{\parindent\hangindent \endgraf \indent \egroup
  \ignorespaces}

\def\item{\varitem to2.5em}
\def\sitem{\varitem to4.5em}
\def\ssitem{\varitem to6.5em}


\newcount\pointn@m   \pointn@m=0
\def\pointbegin{\gz@\pointn@m \point}
\def\point{\g@ne\pointn@m
  \xdef\the@label{\the\pointn@m}\item{\the@label.}}

\newcount\spointn@m   \spointn@m=96
\def\spointbegin{\global\spointn@m96 \spoint}
\def\spoint{\g@ne\spointn@m
  \xdef\the@label{\char\the\spointn@m}\sitem{(\the@label)}}

\newcount\sspointn@m   \sspointn@m=0
\def\sspointbegin{\gz@\sspointn@m \sspoint}
\def\sspoint{\g@ne\sspointn@m
  \xdef\the@label{\romannumeral\sspointn@m}\ssitem{\the@label)}}



\message{matrices and additional math symbols,}


\def\matc{\let\mat@lfil\hfil \let\mat@rfil\hfil}
\def\matl{\let\mat@lfil\relax \let\mat@rfil\hfil}
\def\matr{\let\mat@lfil\hfil \let\mat@rfil\relax}

\def\matrix#1{\null\,\vcenter{\normalbaselines\m@th
    \ialign{$\mat@lfil##\mat@rfil$&&\quad$\mat@lfil##\mat@rfil$\crcr
      \mathstrut\crcr\noalign{\kern-\baselineskip}%
      #1\crcr\mathstrut\crcr\noalign{\kern-\baselineskip}}}\,}

\def\bordermatrix#1{\begingroup \m@th
  \setbox\z@\vbox{%
    \def\cr{\crcr\noalign{\kern2\p@\glet\cr\endline}}%
    \ialign{$##\hfil$\kern2\p@\kern\p@renwd&\thinspace$\mat@lfil##%
      \mat@rfil$&&\quad$\mat@lfil##\mat@rfil$\crcr
      \omit\strut\hfil\crcr\noalign{\kern-\baselineskip}%
      #1\crcr\omit\strut\cr}}%
  \setbox\tw@\vbox{\unvcopy\z@\global\setbox\@ne\lastbox}%
  \setbox\tw@\hbox{\unhbox\@ne\unskip\global\setbox\@ne\lastbox}%
  \setbox\tw@\hbox{$\kern\wd\@ne\kern-\p@renwd\left(\kern-\wd\@ne
    \global\setbox\@ne\vbox{\box\@ne\kern2\p@}%
    \vcenter{\kern-\ht\@ne\unvbox\z@\kern-\baselineskip}\,\right)$}%
  \null\;\vbox{\kern\ht\@ne\box\tw@}\endgroup}


\mathchardef\smallsum=\dq1006
\mathchardef\smallprod=\dq1005

\def\b@mmode{\relax\ifmmode \expandafter\c@mmode \else $\fi}
\def\c@mmode#1\e@mmode{#1}
\def\e@mmode{$}
\def\defmmode#1#2{\def#1{\b@mmode#2\e@mmode}}

\defmmode\{{\lbrace}
\defmmode\}{\rbrace}

\defmmode\,{\mskip\thinmuskip}
\defmmode\>{\mskip\medmuskip}
\defmmode\;{\mskip\thickmuskip}

\defmmode{\Mit#1}{\mit#1}
\defmmode{\Cal#1}{\cal\uppercase\expandafter{#1}}

\def\dotii#1{{\mathop{#1}\limits^{\vbox to -1.4\p@{\kern-2\p@
   \hbox{\tenrm..}\vss}}}}
\def\dotiii#1{{\mathop{#1}\limits^{\vbox to -1.4\p@{\kern-2\p@
   \hbox{\tenrm...}\vss}}}}
\def\dotiv#1{{\mathop{#1}\limits^{\vbox to -1.4\p@{\kern-2\p@
   \hbox{\tenrm....}\vss}}}}

\let\barsymbol -
\mathchardef\tildesymbol=\dq0218
\def\hatsymbol{{\mathchoice{\null}{\null}{\,\,\hbox{\lower 10\p@\hbox
    {$\widehat{\null}$}}}{\,\hbox{\lower 20\p@\hbox
       {$\hat{\null}$}}}}}


\def\begin@stmt{\par\noindent\bpargroup\stmttitlestyle}
\def\adv@stmt#1#2#3#4{\begin@stmt
  \count@\ifx#2#3#1 \else\z@ \glet#2#3\fi \advance\count@\@ne
  \xdef#1{\the\count@}\edef\the@label{#4#1}}

\def\make@stmt{\ \the@label \make@@stmt}
{\catcode`\:\active
  \gdef\make@@stmt{\ \catcode`\:\active \let:\end@stmt}
}
\def\end@stmt{\catcode`\:\@ther \unskip :\stmtstyle
  \enskip \ignorespaces}

\def\defstmt#1#2#3{\expandafter\def@stmt \csname#1\endcsname
  {#2}{#1@stmt@}#3@}
\def\def@stmt#1#2#3#4#5@{\bgroup
  \toks@{\begin@stmt #2\make@@stmt}\toks@ii{#2\make@stmt}%
  \if#4n\xdef#1{\the\toks@}\else
    \xdef#1{\csname#3adv\endcsname \the\toks@ii}%
    \if#4=\edef\n@xt{\expandafter\noexpand
      \csname#5@stmt@adv\endcsname}\else
      \toks@{\empty}\toks@ii\toks@ \if#4c\toks@{\dot@pref}\fi
      \if#4s\toks@{\sect@pref}\if#5c\toks@ii{\sect@dot@pref}\fi \fi
      \if#5a\toks@ii\toks@ \fi
      \edef\n@xt{\noexpand\adv@stmt \csname#3num\endcsname
        \csname#3save\endcsname \the\toks@ \the\toks@ii}\fi
    \expandafter\glet\csname#3adv\endcsname\n@xt \fi
  \egroup}

\def\Prf{\par\noindent\bpargroup\prftitlestyle \case@language\prfhead
  \let\stmtstyle\prfstyle \make@@stmt}

\message{save, restore and start macros,}


\def\save@type{.texsave }  
\newread\test@read
\newwrite\save@write

\def\s@ve{\immediate\write\save@write}
\begingroup \catcode`\:=\active \catcode`\;=\active
  \outer\gdef\save#1 {{\let\,\space
    \immediate\openout\save@write#1\save@type
    \s@ve{;* definitions for :restore #1 \date\space- \thetime\space*}%
    \s@ve{:comment}%
    \s@ve{:mainlanguage:\case@language{german\else english}}%
    \save@page \save@chap
    \save@equ \save@fig \save@tab \save@ref \save@toc \save@foot
    \bgroup \def\n@xt##1.##2{\advance##2\@ne \s@ve{:start##1\the##2}}%
      \n@xt chap.\chapn@m \n@xt sect.\sectn@m
      \n@xt appendix.\appn@m \n@xt equ.\eqn@m
      \n@xt fig.\fign@m \n@xt tab.\tabn@m
      \n@xt ref.\refn@m \n@xt toc.\tocn@m
      \n@xt foot.\footn@m \egroup
    \s@ve{:\ifx\chap@@eq\sect@@eq app\else
      \ifnum\chapn@m=\z@ sect\else chap\fi \fi init}%
    \@save0\@save1\@save2\@save3\@save4%
    \if K\ref@kc \s@ve{:endcomment}\fi
    \save@restore ref \r@names  \save@restore RF \R@names
    \if C\ref@kc \s@ve{:endcomment}\fi
    \save@restore lbl \l@names  \save@restore eq \e@names
    \save@restore fig \f@names  \save@restore tab \t@names
    \s@ve{;*  end of definitions  *}\immediate\closeout\save@write
    }\wlog{* file #1\save@type saved *}}
  \gdef\save@restore#1 {\def\\{\s@ve{:#1restore}%
      \let\\\save@@restore \\}\the}
  \gdef\save@@restore#1{\toks@\expandafter{#1}%
    \s@ve{:dorestore\string#1{\the\toks@}}}
  \gdef\save@page{\s@ve{:\@opt\page@tbn Ttop Bbot Nno *pagenum%
      :\@opt\head@lrac Llef Rrigh Aal Ccen *thead%
      :\@opt\foot@lrac Llef Rrigh Aal Ccen *tfoot%
      :page\@opt\page@ac Aall Cchap *%
      :\@opt\ori@pl Pportrait Llandscape *}%
    \s@ve{:startpage\@opt\page@ac C\the\pageno@pref. *\the\pageno}}
  \gdef\save@chap{\s@ve{:\@opt\chap@page Fno *chappage%
      :\@opt\chap@yn Nno *chapters%
      :\@opt\chap@ar Aarabic Rroman *chapnum}}
  \gdef\save@equ{\s@ve{:equ\@opt\eq@acs Aall Cchap Ssect *%
      :equ\@opt\eq@lrn Lleft Rright Nnone *%
      :equ\@opt\eq@fs Ffull Sshort *}}
  \gdef\save@fig{\s@ve{:\@opt\fig@page Fno *figpage%
      :fig\@opt\fig@acs Aall Cchap Ssect *}}
  \gdef\save@tab{\s@ve{:\@opt\tab@page Fno *tabpage%
      :tab\@opt\tab@acs Aall Cchap Ssect *}}
  \gdef\save@ref{\s@ve{:\@opt\ref@page Fno *refpage%
      :ref\@opt\ref@kc Kkeep Cclear *%
      :ref\@opt\ref@sbn Ssup Bsqb Nnam *%
      :\@opt\yearpage@yp Yyearpage Ppageyear *}}
  \gdef\save@toc{\s@ve{:\@opt\toc@page Fno *tocpage%
      :toc\ifcase\save@@toc \else none\fi}}
  \gdef\save@foot{\s@ve{:\@opt\foot@page Fno *footpage%
      :foot\@opt\foot@be Bbot Eend *%
      :foot\@opt\foot@bp Bsqb Ppar *}}
\endgroup
\def\@opt#1#2#3 #4{\if #1#2#3\fi \if #4*\else
  \expandafter\@opt\expandafter#1\expandafter#4\fi}
\def\save@@toc{\auto@toc chap\or sect}

\newif\ifcr@ss
\begingroup \let\comment=\relax
  \outer\gdef\restore{\@kill0\@kill1\@kill2\@kill3\@kill4%
    {\def\\##1{\glet##1\undefined}%
      \def\n@xt##1{\the##1\global##1\emptyt@ks}%
      \n@xt\l@names \n@xt\e@names \n@xt\f@names \n@xt\t@names
      \n@xt\r@names \let\\\@RFdef \n@xt\R@names}
    \bgroup \let\comment\relax \let\d@rest@re\dorest@re \@restore}
\endgroup
\def\crossrestore#1 {\bgroup \openin\test@read#1\save@type
  \ifeof\test@read \let\n@xt\egroup
    \message{* file #1\save@type missing *}%
  \else \closein\test@read
    \def\n@xt{\@restore#1 }\let\d@rest@re\docr@ss \fi \n@xt}
\def\@restore#1 {\input #1\save@type \egroup}
\def\all@restore{\glet\name@list}
\outer\def\dorestore{\bgroup \catcode`\@\l@tter \num@lett\d@restore}
\def\d@restore#1#2{\egroup \toks@{#2}\d@rest@re#1}
\def\dorest@re#1{\def@name\name@list#1{\the\toks@}}
\def\docr@ss#1{\ifx#1\undefined \cr@sstrue
    \else \expandafter\testcr@ss#1\cr@ss\@@ \fi
  \ifcr@ss \expandafter\testcr@ss\the\toks@\cr@ss\@@ \else \cr@sstrue \fi
  \ifcr@ss\else \expandafter\dorest@re\expandafter#1\fi}
\def\testcr@ss#1\cr@ss#2\@@{\ifx @#2@\toks@{\cr@ss#1}\cr@ssfalse
  \else \cr@sstrue \fi}
\let\cr@ss=\empty

\def\def@name#1#2{\let\name@list#1\add@name#2\xdef#2}
\def\del@name#1{\bgroup
  \def\n@xt##1\\#1##2\\#1##3\@@##4{\global##4{##1##2}}%
  \def\del@@name##1{\expandafter\n@xt\the##1\\#1\\#1\@@##1}%
  \del@@name\l@names \del@@name\e@names \del@@name\f@names
  \del@@name\t@names \del@@name\r@names \del@@name\R@names \egroup}
\def\add@name#1{\del@name#1%
  {\global\name@list\expandafter{\the\name@list\\#1}}}

\def\kill{\num@lett\@k@ll}
\def\@k@ll#1{\def\k@ll##1{\ifx##1\k@ll \let\k@ll\relax \else
    \del@name##1\glet##1\undefined \fi \k@ll}\k@ll#1\k@ll}


\outer\def\startchap{\st@rt\chapn@m}
\outer\def\startsect{\st@rt\sectn@m}
\outer\def\startappendix{\st@rt\appn@m}
\outer\def\startequ{\st@rt\eqn@m}
\outer\def\startfig{\st@rt\fign@m}
\outer\def\starttab{\st@rt\tabn@m}
\outer\def\startref{\st@rt\refn@m}
\outer\def\starttoc{\st@rt\tocn@m}
\outer\def\startfoot{\st@rt\footn@m}

\def\st@rt#1{\gm@ne#1 \global\advance#1}

\message{installation dependent parameters,}




\hoffset@corr@p=86mm   \voffset@corr@p=98mm      
\hoffset@corrm@p=-5mm   \voffset@corrm@p=4mm      
\hoffset@corr@l=134mm   \voffset@corr@l=70mm     
\hoffset@corrm@l=-5mm   \voffset@corrm@l=-4mm     

\catcode`\@=12 

\message{and default options.}



\hbadness=2000  

\newsect        

\mainlanguage\english 

\botpagenum     
\centhead       
\centfoot       
\pageall        
\portrait       
\titlepage      
\nochappage     
\arabicchapnum  
\chapters       
\equchap        
\equshort       
\equright       
\storelist      
\figall         
\figpage        
\nographics     
\taball         
\tabpage        
\refsup         
\refpage        
\refkeep        
\RFlist         
\yearpage       
\tocpage        
\tocnone        
\footsqb        
\footbot        
\footpage       
\matc           

\wlog{summary of allocations:}
\wlog{last count=\number\count10 }
\wlog{last dimen=\number\count11 }
\wlog{last skip=\number\count12 }
\wlog{last muskip=\number\count13 }
\wlog{last box=\number\count14 }
\wlog{last toks=\number\count15 }
\wlog{last read=\number\count16 }
\wlog{last write=\number\count17 }
\wlog{last fam=\number\count18 }
\wlog{last language=\number\count19 }
\wlog{last insert=\number\count20 }

   \def\Fmtversion{2.0}

\edef\fmtversion{\Fmtversion(\fmtname\space\fmtversion)}
\let\fmtname=\Fmtname 
\everyjob={
  \immediate\write16{\fmtname\space version \fmtversion\space
    format preloaded.}%
  \input phystime      
  \input physupdt }    
\immediate\write16{Version \fmtversion\space format loaded.}%


\def\wlog#1{} 
\catcode`\@=11


\def\({\relax\ifmmode[\else$[$\nobreak\hskip.3em\fi}
\def\){\relax\ifmmode]\else\nobreak\hskip.2em$]$\fi}

\def\gappr{\mathpalette\under@rel{>\approx}}
\def\lappr{\mathpalette\under@rel{<\approx}}
\def\gsim{\mathpalette\under@rel{>\sim}}
\def\lsim{\mathpalette\under@rel{<\sim}}
\def\under@rel#1#2{\under@@rel#1#2}

\def\under@@rel#1#2#3{\mathrel{\mathop{#1#2}\limits_{#1#3}}}

\def\under@@rel#1#2#3{\mathrel{\vcenter{\hbox{$%
  \lower3.8pt\hbox{$#1#2$}\atop{\raise1.8pt\hbox{$#1#3$}}%
  $}}}}

\def\widebar#1{\mkern1.5mu\overline{\mkern-1.5mu#1\mkern-1.mu}\mkern1.mu}
\def\parenbar{\mathpalette\p@renb@r}
\def\p@renb@r#1#2{\vbox{%
  \ifx#1\scriptscriptstyle \dimen@.7em\dimen@ii.2em\else
  \ifx#1\scriptstyle \dimen@.8em\dimen@ii.25em\else
  \dimen@1em\dimen@ii.4em\fi\fi \offinterlineskip
  \ialign{\hfill##\hfill\cr
    \vbox{\hrule width\dimen@ii}\cr
    \noalign{\vskip-.3ex}%
    \hbox to\dimen@{$\mathchar300\hfil\mathchar301$}\cr
    \noalign{\vskip-.3ex}%
    $#1#2$\cr}}}


\def\mppae@text{{Max-Planck-Institut f\"ur Physik}}
\def\mppwh@text{{Werner-Heisenberg-Institut}}

\def\mppaddresstext{Postfach 40 12 12, D-8000 M\"unchen 40\else
  P.O.Box 40 12 12, Munich (Fed.^^>Rep.^^>Germany)}

\def\mppaddress{\address{\mppae@text \nl -- \mppwh@text\space --\nl
  \case@language\mppaddresstext}}

\def\mppnum#1{\topright{MPI-Ph/#1}}


\font\fourteenssb=cmssdc10 scaled \magstep2 
\font\seventeenssb=cmssdc10 scaled \magstep3 

\def\letter#1#2{\b@lett@r{26}%
  \centerline{\seventeenssb \uppercase\mppae@text}%
  \centerline{\fourteenssb \uppercase\mppwh@text}%
  \centerline{\strut#1}\vskip.5cm%
  \e@lett@r{\hss\vtop to5cm{\hsize55mm%
    \lftline{\strut}\eightrm  \setbaselineskip=12pt \vfil
    \lftline{F\"OHRINGER RING 6}\lftline{\tenrm D-8000 M\"UNCHEN 40}%
    \lftline{\case@language{TELEFON\else PHONE}: (089) 3 23 08
      \if!#2!\else - #2 \case@language{oder\else or} \fi-1}%
    \lftline{TELEGRAMM:}\lftline{PHYSIKPLANCK M\"UNCHEN}%
    \lftline{TELEX: 5 21 56 19 mppa d}%
    \lftline{TELEFAX: (089) 3 22 67 04}%
    \lftline{POSTFACH 40 12 12}
    \ifx\EARN\undefined\else\vskip5\p@\lftline{EARN/BITNET: \EARN
      @DM0MPI11}\fi \vfil}}}

\def\b@lett@r#1{\endpage \begingroup \doublespace \vglue-#1mm}

\def\e@lett@r#1#2{\skippagenum T\skipheadline T\skipfootline T%
  \line{\vtop to47mm{\lftline{\llap{\vbox to\z@{\vskip171\p@
      \hrule\@width7\p@\vss}\hskip57\p@}\strut}\vskip2mm\vfil
    \addressspacing \dimen@\baselineskip \dimen@ii-2.79ex%
    \advance\dimen@ii\dimen@ \baselineskip\dimen@\@minus\dimen@ii
    \let\nl\cr\use@nl \halign{##\hfil\crcr#2\crcr}\vfil}#1}%
  \vskip1cm\rtline{\thedate}\vskip1cm\@plus1cm\@minus.5cm\endgroup}

\let\addressspacing=\empty


\def\myname{Dr.\ Xxxx Xxxxxxxxxx\nl Physiker}
\def\myaddress{Xxxxxxx Stra\ss e  ??\nl
    \llap{D--8000\quad}M\"unchen ??\nl
    Tel:\ (089) \vtop{\hbox{?? ?? ?? (privat)}%
                      \hbox{3 18 93-??? (B\"uro)}}}
\def\myletter{\b@lett@r{26}\line{\let\nl\cr \use@nl \caps
  \vtop to25mm{\halign{\strut##\hfil\crcr\myname\crcr}\vfil}\hfil
  \vtop to25mm{\tenpoint
    \halign{\strut##\hfil\crcr\myaddress\crcr}\vfil}}%
  \e@lett@r\empty}


\def\firstpageoutput{\physoutput
  \global\output{\setbox\z@\box@cclv \deadcycles\z@}}


\def\veq{\afterassignment\v@eq \dimen@}
\def\v@eq{$$\vcenter to\dimen@{}$$}

\def\veqn{\afterassignment\v@eqn \dimen@}
\def\v@eqn{$$\vcenter to\dimen@{}\eqn$$}

\def\heq{\afterassignment\h@eq \dimen@}
\def\h@eq{$\hbox to\dimen@{}$ }

\def\wlog{\immediate\write\m@ne} 
\catcode`\@=12 

\def\wlog#1{} 
\catcode`\@=11

\outer\def\pthnum#1{\errmessage{***** \string\pthnum\space is no longer
         supported, use \string\mppnum\space instead}}

\def\wlog{\immediate\write\m@ne} 
\catcode`\@=12 



    \equfull
  \footpar \refsqb
\def\eqabbr{Gl\else eq}
\def\eqsabbr{Gln\else eqs}
\def\tolimit_#1{\mathrel{\mathop{\longrightarrow}\limits_{#1}}}
\def\tohoch^#1{\mathrel{\mathop{\longrightarrow}\limits^{#1}}}

\def\N{{\cal N}}
\def\C{{\cal C}}
\def\l{\lambda}
\def\a{\alpha}
\def\b{\beta}
\def\d{\delta}

\def\o{\omega}
\def\dem{\d_\o}
\def\p{\partial}
\def\pmu{\p_\mu}
\def\pmo{\p^\mu}
\def\pnu{\p_\nu}

\def\r{\rho}

\def\uvi{\underline{\varphi}}
\def\vi{\varphi}
\def\ue{u^{(1)}}
\def\Am{A_\mu}
\def\An{A_\nu}
\def\Fmnu{F_{\mu\nu}}
\def\Fmno{F^{\mu\nu}}
\def\Ga{\Gamma}

\def\Gacl{\Gamma_{cl}}
\def\Gagf{\Gamma_{{\rm g.f.}}}
\def\Gainv{\Gamma_{\rm inv}}
\def\pvi{\partial\varphi}

\def\mn{\mu\nu}
\def\v4{\varphi^4}

\def\Gmn{\Ga_{{\mn}}}

\def\ha{{1\over 2}}
\def\g4{(-g)^{{1\over 4}}}
\def\ifa4{{1\over 4!}}
\def\dalam{{\hbox{\frame{6pt}{6pt}{0pt}}\,}}

\def\N{{\cal N}}
\def\C{{\cal C}}
\def\ga{\gamma}

\def\ue{u^{(1)}}

\def\zze{\sqrt{{z_2\over z_1}}}
\def\zez{\sqrt{{z_1\over z_2}}}

\def \frac#1#2 {\hbox{${#1\over #2}$}}
\def\smdm {\underline m \p _{\underline m}}
 \def\tsmdm {\underline m \tilde \p _{\underline m}}
\RF\BRSH{{\caps C. Becchi, A. Rouet, R. Stora},
       {\sl Comm. Math. Phys.} {\bf 42} (1975) 127}
\RF\EK2{{\caps E. Kraus}, {\sl
       The structure of the invariant charge in massive theories
       with one coupling},
 BUTP 93-26, to be published in {\sl Ann.\ of Phys.}}
\RF\EK3{{\caps E. Kraus},
         {\sl Helv.\ Phys.\ Acta} {\bf 67} (1994) 424}
\RF\AOKI{{\caps K.I. Aoki, Z. Hioki, R. Kawabe, M. Konuma, T. Muta},
       {\sl Suppl. Prog. Theor. Phys.} {\bf 73} (1982) 1}
\RF\BOEHM{{\caps M. B\"ohm, W. Hollik, H. Spiesberger},
       {\sl Fortschr. Phys.} {\bf 34} (1986) 687}
\RF\JB1{{\caps F. Jegerlehner},
       {\sl Renormalizing the standard model},
 in: Proceedings of the 1990 Theoretical Advanced Study Institute in
 Elementary Particle Physics, Boulder, Colorado, ed. M. Cvetic and P.
        Langacker, 1991, Singapore }
\RF\KNIEHL{{\caps B.\ A.\ Kniehl}, {\sl Status of higher-order
       corrections in the standard electroweak theory},
       KEK-TH-412, Sept.\ 1994}
\RF\BARBIERI{{\caps R.\ Barbieri, M.\ Beccaria, P.\ Ciafaloni,
          G.\ Curci, A.\ Vicer\'e},
       {\sl Phys. Lett.} {\bf B288} (1992) 95}
\RF\DEGRASSI{{\caps G.\ Degrassi, S.\ Fanchiotti, P.\ Gambino},
 {\sl Two-loop next-to-leading $m_t$ corrections to the
      $\rho$ parameter},
       CERN-TH.7180/94}
\RF\DENNER{{\caps A. Denner, S. Dittmaier, G. Weiglein},
       {\sl Nucl.\ Phys.B, Proc.\ Suppl.} {\bf 37B} (1994) 87}
\RF\SHORE{{\caps G. Shore},
       {\sl Ann.\ of Phys.} {\bf 137} (1981) 262}
\RF\EINHORN{{\caps M.B.\ Einhorn, J.\ Wudka},
       {\sl Phys.\ Rev.} {\bf D39} (1989) 2758}

\topright{MPI-Ph/95-11}
\topright{BONN-TH-95-04}
\topright{February 1995}

\title{Rigid invariance as derived from BRS invariance\nl
       The abelian Higgs model}
\bigskip
\centerline{\caps Elisabeth Kraus\footnote{*}{\rm Supported by
                  DFG}}
\centerline{\sl Physikalisches Institut}
\centerline{\sl Universit\"at Bonn}
\centerline{\sl Nu{\ss}allee 12, D-53115 Bonn}
\bigskip
\centerline{\caps Klaus Sibold}
\centerline{\sl Max-Planck-Institut f\"ur Physik}
\centerline{\sl Werner-Heisenberg-Institut}
\centerline{\sl F\"ohringer Ring 6, D-80805 M\"unchen}
\vskip1cm
\abstract
{Consequences of a symmetry, e.g.\ relations amongst Green
functions, are renormalization scheme independently expressed in terms
of a rigid Ward identity. The corresponding local version yields
information on the respective current. In the case of spontaneous
breakdown one has to define the theory via the BRS invariance and
thus to construct rigid and current Ward identity non-trivially in
accordance with it. We performed this construction to all
orders of perturbation theory in the abelian Higgs model as a
prelude to the standard model.
A technical tool of interest in itself is the
use of a doublet of external scalar ``background'' fields. The
Callan-Symanzik equation has an interesting form and follows easily
once the rigid invariance is established.}
\vfill
\endpage

\chap{Introduction}

The precision of the next generation of experiments requires the
calculation of two-loop contributions in the standard model. These
calculations (for a recent review s.\ \quref{\KNIEHL})
 demand also higher precision as far as the more abstract side
of the theory is concerned. Whereas at one-loop the compatibility
of the renormalization scheme with the underlying theory need not
be really discussed, at two loops this becomes much more urgent.
Similar comments concern the rigid invariance. At one-loop the schemes
used in practice are essentially compatible with the rigid invariance.
In addition one has a more or less complete description available
i.e.\ almost all divergent diagrams have been calculated, almost all
interesting processes have been studied: there is not much need of a
Ward identity (WI) which collects and formalizes the content and the
consequences of rigid invariance. At two loops it will probably
be impossible to perform systematic
 renormalization calculations without
{\it explicit\/} use of the rigid invariance. It will serve at least
as an indispensable tool for checking, but probably help even more
in revealing the symmetry relation amongst Green functions and
amplitudes. As an example where a specific WI was of great help
one may look at \quref{\BARBIERI}, whereas in \quref{\DEGRASSI}
a current algebra argument has been used. The latter arises from
a local WI.

With this in mind
we study in the present paper the rigid invariance in
the abelian Higgs model, in
a subsequent paper in the standard model. We formulate the rigid
invariance in terms of a WI to all orders of perturbation theory.
For these purposes  external scalar ``background'' fields
 have to be introduced in order to absorb the breaking by the
`t Hooft gauge fixing.
It turns out, that in higher orders the classical WI is deformed
in a well-defined way according to the normalization conditions, which
one has imposed.
 Our
presentation is essentially scheme independent and relies on the
inductive construction order by order in the perturbation series.
For this reason the classical approximation is treated extensively
since already there the requirements must uniquely determine the desired
quantities. Higher orders are then seen not to change dramatically
the picture. Furthermore
in the abelian model one is immediately able to derive a local WI.

As an application of rigid and local symmetry we construct the
Callan-Symanzik (CS) equation. With its help it
is possible to compute the asymptotic logarithms of the Green functions,
and to find
their relations according to the symmetries in a simple way.
In an appendix we have collected the propagators of the
model for general values of the gauge parameters.

\chap{The construction of the model}

The model comprises a vector field $A_\mu$ and two scalar fields
$\vi_1,\vi_2$ interacting in such a way that $U(1)$
gauge invariance is spontaneously broken. In conventional
normalization we find that
$$\eqalign{
\Ga_{inv} = &\int\biggl(-{1\over 4}\Fmnu\Fmno + (D_\mu\phi)^*
              (D^\mu \phi) + \ha m^2_H\phi^*\phi\cr
       &\phantom{\int\biggl(-{1\over 4}\Fmnu\Fmno}
             -\ha{m^2_H\over m^2} e^2(\phi^*\phi)^2\biggr)\cr}
\EQN\F21$$
where
$$\eqalign{
\Fmnu\equiv\pmu \An-\pnu\Am,\quad &\phantom{D_\mu}
\phi \equiv{1\over\sqrt{2}}
      (\vi_1+v+i\vi_2)\/,\cr
          &D_\mu\phi  \equiv\pmu\phi -ie\Am\phi\/,\cr}
\EQN\F22$$
is invariant under the transformations
$$
\dem\phi = ie\o\phi\/,\qquad \dem\Am =\pmu\o\/.
\EQN\F23$$
Choosing
$$v={m\over e}
\EQN\F24$$
one can convince oneself that the vector field $\Am$ has mass $m$,
the field $\vi_1$ is the Higgs field with mass $m_H$ and the
field $\vi_2$ is the would-be Goldstone field eaten up by $\Am$.
The vacuum expectation values of the scalar fields are zero:
$$
<\vi_1>=0=<\vi_2>.
\EQN\F25$$
Since $\Gamma_{inv}$ is invariant for $x$-dependent $\o$, it is
a fortiori invariant under transformations with constant $\o$
(``rigid'' transformations).

For the calculation of Green functions and higher order corrections
it is necessary to fix the gauge:
$$
\Gamma_{{\rm g.f.}}=\int(\ha\xi B^2+B(\p A+\xi_A m\vi_2))
\EQN\F26$$
$B$ is an invariant auxiliary field
$$
\d_\o B=0\/.
\EQN\F27$$
The 't Hooft term with $\xi_A\not= 0$ has to be introduced
in order to avoid a non-integrable infrared singularity in the
$<\vi_2\vi_2>$ propagator. This 't Hooft
type gauge fixing violates not only the local gauge invariance,
but also the rigid symmetry non-trivially
$$
\d_\o\Gagf =\int(\o\dalam B+\o B\xi_A m(\vi_1+v))\/.
\EQN\F28$$
Hence it is unavoidable to translate local gauge transformations
into BRS transformations by introducing the Faddeev-Popov $(\phi\pi)$
fields $c,\bar c$
$$\vbox{\halign{$#$\hfill&#&$#$\hfill\quad&$#$\hfill&#&$#$\hfill\cr
s\Am &= &\pmu c &s\bar c &=&B \cr
s c  &= &0      &sB      &=&0 \cr
s\vi_1 &= &-ec\vi_2 &s\vi_2 &=&ec(\vi_1 +v)\cr}}
\EQN\F29$$
and to require BRS invariance instead of (broken) gauge
invariance in order to define the theory
$$
\Gacl=\Ga_{inv}+\Gagf +\Ga_{\phi\pi}+\Ga_{{\rm ext.f.}}
\EQN\F210$$
Here
$$
\Ga_{\phi\pi} =\int\Bigl(-\bar c~\dalam c-e\bar c \xi_Am
            (\vi_1 + v) c \Bigr)
\EQN\F210a$$
and we have furthermore added an external field dependent part
$$
\Ga_{{\rm ext.f.}}=\int\Bigl(Y_1(-ec\vi_2)+Y_2(ec(\vi_1+v))\Bigr)
\EQN\F211$$
because the BRS transformations are non-linear in propagating
(and interacting) fields, hence have to be defined in higher
orders in a nontrivial way.

The invariance under BRS transformations can be expressed
in terms of the vertex functional $\Ga$
as the Slavnov-Taylor identity (ST)
$$
s(\Ga)\equiv\int\left(\p c{\d \Ga\over\d A}+B{\d\Ga\over\d\bar c}+
   {\d\Ga\over\d\underline{Y}}\cdot {\d\Ga\over\d\uvi}\right)=0\/.
\EQN\F212$$
$\Gacl$ is the lowest order in the perturbative expansion of $\Ga$,
the tree approximation. In terms of $Z$, the generating functional
of Green functions, the ST identity reads
$$
sZ\equiv\int\left(\pmu J^\mu{\d Z\over\d J_c}-J_{\bar c}
{\d Z\over\d J_B}+J_{\underline{\vi}} {\d Z\over\d \underline{Y}}
\right) =0\/,
\EQN\F213$$
Here $J_x$ with $x=c,\bar c,B,\mu,\vi $ denotes the sources for the
respective fields. In addition to \queq{\F212} $\Gacl$ solves the
gauge condition
$$
{\d\Ga\over\d B} =\xi B+\p A+\xi_Am\vi_2
\EQN\F214$$
which can be imposed in this form to all orders of perturbation
theory.
The ghost equation of motion
$$
{\d\Ga\over\d\bar c}+\xi_A m{\d\Ga\over\d Y_2} =-\dalam c
\EQN\F215$$
follows from \queq{\F212} and \queq{\F214}.

The importance of the ST identity originates on the one hand from
the fact that it permits to prove unitarity, i.e.\ the possibility
of constructing a Hilbert space of physical states within which the
scattering proceeds (for physical initial states). On the other hand --
as alluded to above -- it defines the model in question once multiplets
have been chosen and normalization conditions have been specified.
This is a renormalization scheme independent procedure and therefore
unquestionable, whereas giving an action and its counterterms is not.

In order to see which normalization conditions are needed we
present explicitly the general solution of the ST identity
\queq{\F212} and the gauge condition \queq{\F214} in the tree
approximation. It has the form
$$\eqalignno{
\Gacl^{gen}=&\int(\ha\xi B^2+B(\p A+\xi_A m\vi_2)-\bar c~\dalam
              c) +\hat\Ga   &\EQ\F217\cr
\hat\Ga    =&\Lambda (A,\vi_1,\vi_2)+
      \int \hat e(-Y_1 z_2\vi_2 +Y_2z_1(\vi_1 +v) &\cr
    &\phantom{\Lambda (A,\vi_1,\vi_2)+ \int}
            -\xi_A m\bar c z_1(\vi_1+v))c
                           &\subeqbegin\cr
\Lambda    =&\int\Bigl(-{z_A\over 4}\Fmnu\Fmno +{z_1\over 2} \pvi_1
              \pvi_1+{z_2\over 2}\pvi_2\pvi_2     &\subeq\cr
            &+ \hat e z_1z_2 (\pvi_1\vi_2 -\pvi_2\vi_1) A+\ha
                          \hat e^2z_1z_2(z_1\vi_1^2+z_2\vi_2^2)A^2\cr
            &+\ha \hat e^2z_2z_1^2v^2A^2-z_1z_2 \hat ev\pvi_2 A+
                     \hat   e^2z_1^2z_2v\vi_1 A^2\cr
            &+\ha\mu^2(z_1\vi_1^2 +2z_1v \vi_1+z_2\vi_2^2)-{1\over 4}
   \hat  \lambda (z_1\vi_1^2+2z_1v\vi_1+z_2\vi_2^2+z_1v^2)^2\cr}
$$
For the derivation of \queq{\F217} we have also imposed invariance
under charge conjugation (see table for quantum numbers).
\midinsert
\centerline{Table: {\sl Quantum numbers}}
\bigskip
\centerline{\table to \hsize{
#\hfill&\vrule\hfill#\hfill&\vrule\hfill#\hfill&\vrule\hfill#\hfill
&\vrule\hfill#\hfill&\vrule\hfill#\hfill&\vrule\hfill#\hfill&\vrule
\hfill#\hfill&\vrule\hfill#\hfill&\vrule\hfill#\hfill\cr\hrule
fields &$A_\mu$ &$B$ &$\tilde\varphi_1$ &$\tilde\varphi_2$ &$c$ &$\bar c$
&$Y_1$ &$Y_2$ &$q_1$ \cr\hrule
dim           &1 &2 &1 &1 &0 &2 &3 &3 &1 \cr\hrule
charge conj.\ &--&--&+ &--&--&--&+ &--&+ \cr\hrule
$Q_{\phi\pi}$ &0 &0 &0 &0 &+1 &--1 &--1 &--1 &+1 \cr\hrule
}}
\bigskip \noindent $\tilde\varphi =\varphi, \hat\varphi$
\endinsert
The wave function normalizations $z_1,z_2$ and $z_A$, the masses of
the vector and the Higgs particle, i.e.\ the parameters
 $v, \mu, \hat\lambda$
and the coupling $\hat e$ are not prescribed by the ST identity. They
have to be fixed by appropriate normalization conditions to all
orders.

In order to have a particle interpretation we shall fix
the mass poles for the physical particles:
$$\eqaligntag{
\Ga_{\vi_1\vi_1}(p^2=m_H^2)&=0  &\EQADV{\F219}\SUBEQBEGIN\eqba\cr
\Ga^T(p^2=m^2)              &=0  &\SUBEQ\eqbb\cr}
$$
for $\Gmn =\Bigl(\eta_{\mn}-{p_\mu p_\nu\over p^2}\Bigr)\Ga^T+
                     {p_\mu p_\nu\over p^2}\Ga^L$
and require vanishing vacuum expectation value for $\vi_1$:
$$
<\vi_1> =0 \SUBEQN\eqbc
$$
\queq{\F219 \eqba -\eqbc} fixes the parameters $v, \hat \l$ and $ \mu$.

A complete on-shell version of the model is then, in analogy to
the standard model \quref{\AOKI,\BOEHM,\JB1},
 defined by fixing also their residues there.
These normalization conditions determine $z_1$ and $z_A$:
$$\eqalignno{
\p_{p^2}\Ga_{\vi_1\vi_1}(p^2=m^2_H)&=1    &\SUBEQ\eqbd\cr
\p_{p^2}\Ga^T(p^2=m^2) &=1                &\SUBEQ\eqbe\cr
}$$
Only
 the amplitude $z_2$ of the field $\vi_2$ will be prescribed off-shell
$$
\p_{p^2}\Ga_{\vi_2\vi_2} (p^2=\kappa^2)=1 \SUBEQN\eqbf
$$
and  the coupling $\hat e $ is also fixed at an arbitrary normalization
momentum $\kappa ^2$
$$
\p_p\Ga_{\vi_1\vi_2 A}   (p=p_{sym}(\kappa)) = ie \SUBEQN\eqbg
$$
The gauge parameter $\xi_A$ is introduced by requiring
$$
\Ga_{c\bar c}(p^2=m^2_{{\rm ghost}})=
0\qquad m^2_{{\rm ghost}} =\xi^{(o)}_Am^2
\SUBEQN\eqbh
$$
In higher orders the $\hbar$-contributions of
 $\xi_A=\xi_A^{(o)}+O(\hbar)$
guarantee that indeed a pole is generated at $\xi_A^{(o)}m^2$ for the
ghost propagator.

The model is thus described as in the standard form in terms of the
following physical parameters: vector mass $m$, Higgs mass $m_H$,
charge $e$.

It is instructive (e.g.\ for the derivation of
Callan-Symanzik and renormalization group equation) to go over
from the general to the standard form by an explicit redefinition
of fields and parameters:
$$\vbox{\halign{
$#$\hfill &#&$#$\hfill\qquad\qquad&$#$\hfill &#& &$#$\hfill\cr
\vi_1^o &=&\sqrt{z_1}\vi_1                 
&Y_1^o  &=&{1\over\sqrt{z_1}} Y_1\cr
\vi_2^o &=&\sqrt{z_2}\vi_2                 
&Y^o_2  &=&{1\over\sqrt{z_2}} Y_2\cr
A^o_\mu &=&\sqrt{z_A} \Am                  
&c^o &=&\sqrt{z_A} c\cr
B^o     &=&{1\over \sqrt{z_A}}B                  
&\bar c^o &=& {1\over\sqrt{z_A}}\bar c\cr
v^o &=&\sqrt{z_1} v                        
&\mu^o &=& \mu\cr
\lambda^o &=&\hat \lambda                       
&\xi^o_A &=& {z_A\over\sqrt{z_2}}\xi_A\cr
\xi^o &=& z_A\xi &e^o&=&\sqrt{z_1 z_2 \over z_A} \hat e\cr}} 
\EQN\F220
$$
In a scheme, where renormalized and bare quantities are
distinguished this constitutes their relation.
It can be shown \quref{\BRSH} that ST identity \queq{\F212},
gauge condition \queq{\F214}, charge conjugation invariance and the
normalization conditions \queq{\F219} indeed uniquely define the
model to all orders.
This means that the Green functions of the model are finite,
unambigously specified and independent of the renormalization
scheme one has used in the course of their calculation.
We have displayed this construction of the model, in particular
the general solution and the normalization conditions in such
detail because they are needed for the derivation of the rigid
invariance from BRS invariance.

\chap{Rigid invariance}

A symmetry relates a priori unrelated Green functions and gives
thus rise to observable consequences like grouping particles
into multiplets or relations amongst scattering amplitudes.
 Usually these informations are contained in
Ward-identities (WI) or in conservation equations for currents
and charges. We were forced to define our model via the ST identity
and have therefore now the task to derive these relations in accordance
with the latter. In the present section we shall prove a linear WI,
in the next section the current conservation equation.

\sect{Classical approximation}

It is obvious that rigid invariance of $\Gainv$ \queq{\F21} can be
expressed by the WI
$$
W\Gainv\equiv\int\Bigl(-\vi_2{\d\over\d\vi_1}+(\vi_1+v){\d\over\d\vi_2}
\Bigr)\Gainv =0
\EQN\F31$$
For $\Gacl$ \queq{\F210}  one finds immediately, that the WI is broken by
the t'Hooft gauge fixing:
$$\eqalign{
W\Gacl\equiv &\int\left(-\vi_2{\d\over\d\vi_1}+(\vi_1+v)
{\d\over\d\vi_2}-Y_2{\d\over\d Y_1}+Y_1{\d\over Y_2}\right)\Gacl\cr
= &\int\left(\xi_A mB(\vi_1+v)+\xi_Ame\bar c c\vi_2\right)\cr
= &s\int\xi_A m\bar c (\vi_1+v)\cr}
\EQN\F32$$
 here $v={m\over e}$.
More interesting, namely pointing into the direction of our present
problem, is the {\it postulate\/} requiring rigid invariance on the
general classical action \queq{\F217} up to the t'Hooft breaking:
$$
W\Gacl^{{\rm gen}} =s\int\xi_A m\bar c(\vi_1+{m\over e})
\EQN\F33$$
because it enforces the relation
$$
z_1=z_2 
\EQN\F34$$
A rigid invariance as specified by  \queq{\F33} replaces
 one normalization condition, it relates
 the normalization of the wave function
$z_2$ to the normalization $z_1.$
But if one wants to  calculate Green functions in the on-shell
normalization, one has to ensure the desired normalization
by an explicit normalization condition. Therefore we allow an
appropriate deformation of the WI operator rather than using the
WI as normalization condition, i.e.\ we stick to \queq{\F219}
and study the consequences for the WI.

In the tree approximation it can be read off from the ST identity
for constant ghost fields that the following general WI holds
$$\eqalign{
W^{{\rm gen}}\Gacl^{{\rm gen}}\equiv &z\int\biggl(-
\sqrt{{z_2\over z_1}}\vi_2{\d\over \d \vi_1}
+\sqrt{{z_1\over z_2}} (\vi_1+v)
      {\d\over\d\vi_2}\cr
&\quad -\sqrt{{z_1\over z_2}} Y_2 {\d\over\d Y_1} +
\sqrt{{z_2\over z_1}}Y_1{\d\over\d Y_2}\biggr)\Gacl^{{\rm gen}}\cr
=& z\sqrt{{z_1\over z_2}} s\int\xi_A m\bar c (\vi_1 +v)\cr}
\EQN\F35$$
(The factor $z$ indicates that the overall normalization is
arbitrary.) $W^{{\rm gen}}$ indeed qualifies for a WI operator
in this abelian model:
it is odd under charge conjugation. In this sense
it is a deformed version of $W$ in \queq{\F32} and reduces
to it in the tree approximation
when the normalization conditions \queq{\F219} are applied.
The factors $z_1,z_2$ will get their real content in higher orders,
but serve here as indicator of what type of deformation is at least
to be expected there.

\sect{Classical approximation with external fields}

By now the calculations of rigid invariance \queq{\F32},
\queq{\F35} have been
carried out in the classical approximation where the vertex
functional is a completely local object. In higher orders the possible
forms of rigid invariance cannot be easily read off from the
ST identity since the ghost fields $c$ and $\bar c$ interact,
in particular appear in the internal lines of the loop corrections.
For the same reason the r.h.s.\ of the WI -- the breaking by the
't~Hooft gauge fixing -- becomes a true insertion and requires a
non-trivial and unambigous definition.

In order to proceed we couple the breaking of rigid
invariance in the classical approximation to the action by
introducing suitably transforming external fields
\quref{\SHORE, \EINHORN, \DENNER}. The
aim is to render the rigid WI homogeneous, hence a
doublet $\left({\hat\vi_1}\atop{\hat\vi_2}\right)$
of scalar external fields is appropriate, since the
breaking transforms as a doublet. BRS invariance is
not broken by the gauge fixing, so there is some freedom
in choosing the BRS transformation properties of $\hat{\uvi}$.
Since the breaking in the rigid WI is a BRS variation we can
prescribe that $\hat{\uvi}$ transforms under BRS as a doublet too
$$
s\pmatrix{\hat\vi_1 \cr \hat\vi_2\cr} =
           \pmatrix{q_1\cr q_2\cr}\/.
\EQN\F36$$
This assignment will also turn out to be very natural when we
study the gauge parameter dependence of the theory in an
algebraic way. Under charge conjugation we require $\hat\vi_1, q_1$,
to be even, $\hat\vi_2, q_2$ to be odd. The (ultraviolet) dimension
of all of them is taken to be 1.

The ST identity is enlarged to
$$
s(\Ga)\equiv\int\left(\pmu c{\d\Ga\over\d A}+ B{\d\Ga\over\d\bar c}
    +{\d\Ga\over\d\underline{Y}}
    \cdot{\d\Ga\over\d\uvi}+\underline{q}
    {\d\Ga\over\d\hat{\uvi}}\right) = 0
\EQN\F37$$
and has to be solved in this generality.
The general solution $\Gacl^{{\rm gen}}$
can be decomposed quite analogously to \queq{\F210} and \queq{\F217} as
$$\Gacl^{{\rm gen}}=\Lambda (A,\bar\vi_1,\bar\vi_2)+\Gagf +
\Ga_{\phi\pi}+\Ga_{{\rm ext.f.}}
\EQN\F38$$
where
$$
\bar\vi_1\equiv\vi_1 -x_1\hat\vi_1\quad ,\quad
\bar\vi_2\equiv\vi_2 -x_2\hat\vi_2\/.
\EQN\F39$$
$\Lambda$ is given in \queq{\F217 b}, with $\vi\to\bar\vi $.
$$
\eqalign{\Ga_{{\rm ext.f.}}=&\int (Y_1(-ez_2\bar\vi_2 c+x_1q_1)\cr
    &\quad +Y_2(e z_1(\bar\vi_1+ v)c +x_2q_2)\cr}
\EQN\F310$$
$x_{1,2}$ are new free parameters of the model. Therefore
in addition to \queq{\F219}  we have to give normalization conditions
for the parameters $x_1 $ and $ x_ 2$ which we choose on the
external field part:
$$\eqaligntag{
\Ga_{Y_2q_2\big|p^2=\kappa^2} &= x_2&\EQADV\F337\subeqbegin\cr
\Ga_{Y_1q_1\big|p^2=\kappa^2} &=x_1 &\subeq\cr}
$$
The gauge fixing terms are not restricted by the ST identity so we take
a linear gauge in the propagating fields:
$$
\Gagf =\int\left(\ha\xi B^2+B\p A-eB\Bigl( (\hat\vi_1-\xi_A {m\over e})
    \vi_2-\hat\vi_2(\vi_1-\hat\xi_A {m\over e})\Bigr)\right)
\EQN\F311$$
The contribution $\xi_AmB\vi_2$ was the starting point ('t~Hooft
gauge), its rigid variation is coupled to $\hat\vi_2$, thus
maintaining charge conjugation invariance; the term
$B\hat\vi_1\vi_2$ is put in for having all linear terms;
factors are chosen for later convenience. The parameter $\xi_A$
is fixed  by \queq{\F219\eqbh}, $\hat\xi_A$
is a further free parameter, which is seen to be fixed by the WI
below (cf.~$(3.18)$).
Terms bilinear in the external fields are omitted because they will
not be created by radiative corrections.
To be more explicit: the gauge condition
$$
{\d\Ga\over\d B}=\xi B+\p A - e\left(\Bigl(
(\hat\vi_1-\xi_A{m\over e}\Bigr)\vi_2 -\hat\vi_2
\Bigl(\vi_1-\hat\xi_A {m\over e}\Bigr)\right)
\EQN\F312$$
can be postulated and integrated to yield $\Gagf$ in all
orders. This means in particular that the normalization of
the fields $\hat\vi_{1,2}$ is implicitly fixed by \queq{\F311}.

Once $\Gagf$ is given, the $\phi\pi$-part is prescribed by the
ST identity:
$$\eqalign{
\Ga_{\phi\pi}=&\int\Bigl(-\bar c~\dalam c+e\bar c(q_1\vi_2 -q_2
     (\vi_1-\hat\xi_A{m\over e}))\cr
&\quad +e\bar c(\hat\vi_1-\xi_A{m\over e})(ez_1(\bar\vi_1+ v)
     c+x_2q_2)\cr
&\quad -e\bar c\hat\vi_2(-ez_2\bar\vi_2 c+x_1q_1)\Bigr)\cr}
\EQN\F313$$
The ghost equ.\ of motion has the form
$$\eqalign{
{\d\Ga\over\d\bar c}+e\bar\vi_2 {\d\Ga\over\d Y_1}-e&(\hat\vi_1-\xi_A
    {m\over e}){\d\Ga\over\d Y_2}\cr
&=-\dalam~c +eq_1\vi_2 -eq_2\Bigl(\vi_1-\hat\xi_A {m\over e}\Bigr)\cr}
\EQN\F314$$

Returning now to the discussion of rigid invariance we first note
that from $\Lambda$ and $\Ga_{{\rm ext.f.}}$
a separate transformation law for $\vi$ and $\hat\vi$ cannot
yet be derived, since they only depend on the combination $\bar\vi$.
But when requiring a rigid invariance on $\Gagf$ and on
 $\Ga_{\phi\pi}$,
then separate transformation laws for $\vi$ and $\hat\vi$ emerge
and one
 is led to the invariance of the general classical action \queq{\F38}
in the following form
$$\eqalign{
& W^{\rm gen}\Gacl^{\rm gen}  \cr
& \equiv \int\biggl(-\sqrt{\frac {z_2}{ z_1} }
\vi_2{\d\over\d\vi_1}
+\sqrt{\frac {z_1}{ z_2} }(\vi_1 - \hat \xi_A \frac me ){\d\over\d\vi_2}-
\sqrt{\frac {z_1}{ z_2} } Y_2{\d\over\d Y_1}+\sqrt{\frac {z_2}{ z_1} }
   Y_1{\d\over\d Y_2}\cr
& \phantom{\equiv \int \,}
 -\sqrt{\frac {z_2}{ z_1} }\hat\vi_2 {\d\over\d\hat\vi_1}+
   \sqrt{\frac {z_1}{ z_2} }(\hat\vi_1-\xi_A \frac me  )
   {\d\over\d\hat\vi_2}-\sqrt{\frac {z_2}{ z_1} }q_2{\d\over\d q_1}
+\sqrt{\frac {z_1}{z_2} } q_1{\d\over\d q_2}\biggr)\Gacl ^{\rm gen}=0\cr}
\EQN\F316$$
This WI restricts the parameters $x_1$ and $x_2$,
in the classical approximation one has
$$
x_1=x_2\equiv x .
\EQN\F317$$
 Also the free parameter
  $\hat \xi _A $ is determined by the WI \queq{\F316}
in terms of $v, x$ and $\xi_ A$
$$
-\hat\xi_A\frac me =  v - x\xi_A{m\over e}
\EQN\F318$$
If one applies the normalization conditions on the classical action,
then
 $z_1=z_2=1$, but these factors indicate,
how the classical transformations may be deformed in higher orders, if we
fix $z_1$ and $z _2$ by independent normalization conditions as
e.g.~\queq{\F219\eqbd} and \queq{\eqbf}.
The tree value of $\hat \xi _A^{(o)}$ is given by
$\hat\xi_A ^{(o)} = -1 + x \xi_A $.

It is easily verified that the WI \queq{\F316} reproduces the breaking
\queq{\F32} at  $\hat{\uvi} =0$ due to the inhomogeneous
term $-\sqrt{{z_1\over z_2}}\xi_A{m\over e}{\d\over\d\hat\vi_2}$
in the WI operator:
$$\eqalign{
&\int\left(-\sqrt{{z_2\over z_1}}\vi_2{\d\over\d\vi_1}+
  \sqrt{{z_1\over z_2}}(\vi_1+v){\d\over\d\vi_2}-\sqrt{{z_1\over z_2}}
  Y_2{\d\over\d Y_1}+\sqrt{{z_2\over z_1}} Y_1
{\d\over\d Y_2}\right)\Gacl
             ^{\rm gen} \Big|_{{\hat{\uvi}=0=\underline{q}}} \cr
&\qquad\qquad =
\sqrt{{z_1\over z_2}}\xi_A{m\over e} \int  \Bigl( {\d
\Gacl ^{\rm gen}   \over\d\hat\vi_2} -x
{\d \Gacl ^{\rm gen}   \over\d\vi_2} \Bigr)
\Big|_{{\hat{\uvi}=0=\underline{q}}}\cr
&\qquad\qquad =\sqrt{{z_1\over z_2}}\int\left(B(\vi_1+v)+z_2e
\bar c\vi_2 c\right)\xi_A m\cr}
\EQN\F319$$
This concludes our presentation of the classical approximation.

\sect{Higher orders}
The first task when looking into higher orders is to establish
the enlarged ST identity \queq{\F37} to all orders. As compared
to \queq{\F212} this requires solving the cohomology problem with
doublet $(\hat{\uvi},\underline q)$ included. We do not reproduce
the respective calculations here but just note, that this cohomology
is trivial hence \queq{\F37} holds once suitable counterterms
are admitted.

Our real aim is to demonstrate the validity of the deformed WI
\queq{\F316} to all orders, when acting on the
 generating functional of 1PI Green functions $\Ga$.
 The most important ingredient for the proof is to note that
$W^{\rm gen}$ has symmetry
 properties with respect to BRS invariance. This
means the following: The action principle tells us that
$$
W^{\rm gen}\Ga =\Delta\cdot\Ga
\EQN\F321$$
where $\Delta$ is a local integrated insertion (i.e.\
a sum of integrated field monomials) with dimension less than
or equal to four, odd under charge conjugation and $\phi\pi$-charge
zero.
 Acting now with $W^{\rm gen}$
 on the ST identity
\queq{\F37} we obtain
$$\eqaligntag{
0=&W^{\rm gen} s(\Ga) = s_\Ga (W ^{\rm gen} \Ga)= s_\Ga (\Delta\cdot\Ga)
               &\EQADV\F322\subeqbegin\cr
\noalign{where}\cr
s_\Ga \equiv&\int\left(\p c{\d\over\d A}+B{\d\over\d\bar c}
+{\d\Ga\over\d\underline{Y}}\cdot{\d\over\d\uvi}+{\d\Ga\over\d\uvi}
\cdot{\d\over\d\underline{Y}}+
\underline{q}{\d\over\d\hat{\vi}}\right)\/.
               &\subeq\cr}
$$
i.e.~$\Delta$ is BRS invariant. Hence we call the differential operator
$W^{\rm gen}$ BRS symmetric.

We derive the validity of a rigid WI by induction starting from the tree
approximation,
 where we have verified (cf.~\queq{\F316} with $z_1 = z_2 =1$)
$$
W\Gacl =0
\EQN\F322A
$$
$W$ is the usual WI operator of rigid invariance as given in \queq\F32 ,
including the external fields $\hat \vi_{1,2} $ and $q _{1,2}$.
{}From \queq{\F322A} follows that $\Delta$ is of order $\hbar$
$$\Bigl( W \Ga \Bigr)^{(\le 1)} = \Delta ^{(1)}
\EQN\F322B
$$
and therefore:
$$
(s_\Ga(\Delta\cdot\Ga))^{(1)}=s_{\Gacl}\Delta ^{(1)} = 0\/,
\EQN\F323$$
\queq{\F323} constitutes a consistency condition for $\Delta$.
Solving it is again solving a cohomology problem,
now in the sector defined by the quantum
numbers of $\Delta$ (charge conjugation: --, $Q_{\phi\pi}:0$).
It turns out that the cohomology is trivial i.e.\
$$
\Delta =s_{\Gacl}\hat\Delta
\EQN\F324$$
$(s_{\Gacl}$ is nilpotent: $s_{\Gacl} s_{\Gacl}=0$), hence the list
of all $\Delta$ is fairly short.
$$\eqalign{
\{\Delta_i\}=s_{\Gacl}\int &
       Y_2\vi_1,Y_1\vi_2,Y_2,Y_2\hat\vi_1,Y_1\hat\vi_2\/,\cr
&\bar c\vi_1,\bar c\hat\vi_1,
\bar c\vi_1\hat\vi_1,\bar c\vi_2\hat\vi_2,\cr
&\bar c, \bar c\hat\vi_1^2,\bar c\hat\vi^2_2,
\bar c\vi^2_1,\bar c\vi^2_2,
    \bar c A^2\cr}
\EQN\F325$$
A glance on the terms containing $\bar c$ shows that those of the
first line were ``used'' for the gauge fixing \queq{\F311}, wheras
those of the second line were not ``used''. This will soon be seen to
be relevant for the coefficients with which they appear in \queq{\F321}.
In order to determine them we rewrite the
monomials as differential operators to the extent to which this is
possible.
For some this is obvious:
$$\eqalign{
s_{\Gacl}\int Y_2\vi_1 = &\int\left(\vi_1{\d\Gacl\over\d\vi_2} -
   Y_2{\d\Gacl\over\d Y_1}\right)\cr
s_{\Gacl}\int Y_1\vi_2 = &\int\left(\vi_2{\d\Gacl\over\d\vi_1} -
   Y_1{\d\Gacl\over\d Y_2}\right)\cr
s_{\Gacl}\int Y_2 = &\int {\d\Gacl\over\d\vi_2}\cr
s_{\Gacl}\int Y_2\hat\vi_1=&\int\left(\hat\vi_1 {\d\Gacl\over\d\vi_2} -
   Y_2 q_1\right)\cr
s_{\Gacl}\int Y_1\hat\vi_2 =&\int\left(\hat\vi_2{\d\Gacl\over\d\vi_1}-
   Y_1q_2\right)\cr}
\EQN\F326$$
For some others this rewriting requires a little calculation:
$$\eqalign{
\int {\d\Gacl\over\d\hat\vi_2} = &s_{\Gacl}\int(-x Y_2+e\bar c(\vi_1 -
     \hat\xi_A {m\over e}))\cr
\int\left(\hat\vi_2{\d\over\d\hat\vi_1}+q_2{\d\over\d q_1}\right)
     \Gacl =&s_{\Gacl}\int(-x \hat\vi_2 Y_1-e\bar c\hat\vi_2\vi_2)\cr
\int\left(\hat\vi_1{\d\over\d\hat\vi_2} +
q_1{\d\over\d q_2}\right)\Gacl = &s_{\Gacl}\int
    \left(-x \hat\vi_1 Y_2+e\bar c\hat\vi_1(\vi_1-\hat\xi_A
    {m\over e})\right)\cr}
\EQN\F228$$
{}From this representation it is also clear that these differential
operators are symmetric with respect to BRS like $W^{\rm gen}$, in fact they
constitute just $W^{\rm gen}$! The validity of the classical WI \queq{\F322A}
 implies immediately that not all of the
above differential operators are linearly
independent when acting on $\Gacl$ and that consequently the polynomials in
\queq{\F325} are not independent.
Hence we have to eliminate one of the polynomials in
the first line of \queq\F325 \ e.g.~$s_{\Gacl}\int(Y_1\vi_2)$.
Now we are able to  rewrite \queq\F322B \ as follows
$$\eqalign{
(W\Ga)^{(\leq 1)} =  \int\biggl(& -
\bigl(\ue_1 \vi_1+ v^{(1)}\bigr) {\d\Ga\over\d\vi_2}
 +\ue_1 Y_2{\d\Ga\over\d Y_1}\cr
& + \ue_3\bigl(\hat\vi_2 {\d\Ga\over\d\hat\vi_1}+  q_2{\d\Ga\over\d q_1}
 \bigr) \cr
&-\bigl(\ue_4\hat\vi_1 + w ^{(1)} \bigr)
{\d\Ga\over\d\hat\vi_2}
-\ue_4 q_1{\d\Ga\over\d q_2}
\biggr)\cr
 + \biggl(&  \ue_5\Bigl(\hat\vi_1{\d\Ga\over\d\vi_2}-Y_2q_1\Bigr)+
\ue_6\Bigl(\hat\vi_2{\d\Ga\over\d \vi_1}-Y_1q_2\Bigr)\cr
& +s_{\Gacl}\Bigl(\ue_7\bar c +\ue_8\bar c\hat\vi_1+\ue_9\bar c
\hat\vi^2_1+\ue_{10}\bar c\hat\vi^2_2\cr
&\phantom{+s_{\Gacl}} \, +\ue_{11}\bar c\vi^2_1+\ue_{12}\bar c\vi^2_2+\ue_{13}
\bar c A^2\Bigr)\biggr)\cr}
\EQN\F328$$
The test on the gauge fixing condition \queq\F312 \ leads to the following
relations among the coefficients:
$$\eqalign{
\ue_4 = \ue_1 \/,\, \ue _3 = 0 &\qquad \ue_{11}=\ue_{12}=\ue_{13}=0\/, \cr
\ue_9 = e \ue_5 &\quad \ue_{10} = -e \ue _6}
\EQN\F329$$
By appropriate choice of the shift parameter
and by fixing thereby implicitly
$\hat\xi_A$
we can rewrite \queq\F328 \ in the following form $ (\ue_1 \equiv \ue)$:
$$\eqalign{
\Bigl((W+ \d W ^{(1)}) \Ga\Bigr)^{(\leq 1)}\equiv \int\biggl(&
-\vi_2{\d \over\d\vi_1}+\Bigl((1+\ue )(\vi_1
                        -\hat \xi _A \frac me )\Bigr)
{\d \over\d\vi_2}\cr
 & +  Y_1 {\d \over\d Y_2} -(1+\ue ) Y_2{\d \over\d Y_1}\cr
  & -  \hat\vi_2{\d \over\d\hat\vi_1}+\Bigl(
(1+\ue )(\hat\vi_1-\xi_A \frac me )
{\d \over\d\hat\vi_2}\cr
  & - q_2{\d\over\d q_1}+(1+\ue )q_1{\d\over\d q_2}
\biggr)\Ga\cr
 =\int\biggl(&\ue_5\Bigl(\hat\vi_1{\d\Gacl \over\d\vi_2}-Y_2q_1\Bigr)+
\ue_6\Bigl(\hat\vi_2{\d\Gacl \over\d \vi_1}-Y_1q_2\Bigr)\cr
&\quad +s_{\Gacl}\Bigl(-\xi_A m\ue_5\bar c\hat\vi_1+e\ue_5
\bar c\hat\vi^2_1-e\ue_{6}\bar c\hat\vi^2_2\Bigr)\biggr)\cr}
\EQN\F330$$
Thereby we have taken all operators which appear already in the classical
WI operator  on the l.h.s.\ defining a deformed WI operator
in 1-loop order
$$W_1 = W + \d W^{(1)}\EQN\F330A$$
$W_1$ is quite  analogous to the operator $W^{\rm gen}$ \queq\F316 \
 of the classical approximation.

It remains to be shown,
 that the r.h.s.\ of \queq\F330\ is actually vanishing
in 1-loop order.
Testing with respect to $Y_2q_1$ resp.\ $Y_1q_2$ yields equations
for $\ue_5$ and $\ue_6$:
$$\eqaligntag{
\ue_5=&-\Ga_{Y_1q_1}^{(1)}+ \Ga_{Y_2q_2}^{(1)}
                -\hat\xi_A \frac me \Ga_{\vi_2Y_2q_1}^{(1)}
   -\xi_A\frac me \Ga_{{\hat\vi_2}Y_2q_1}^{(1)}   &\EQ\F331\cr
\ue_6=&-\Ga_{Y_1q_1}^{(1)} +\Ga_{Y_2q_2}^{(1)}
                -\hat\xi_A \frac me \Ga_{\vi_2Y_1q_2}^{(1)}
   -\xi_A \frac me \Ga_{{\hat\vi_2}Y_1q_2}^{(1)}  &\EQ\F332\cr}
$$
The three-point-functions disappear in the limit of infinite momentum,
hence
$$
\ue_5 =\ue_6\/.
\EQN\F333$$
The WI becomes
$$\eqalign{
(W_1\Ga)^{(\leq 1)}=
       \ue_5 & \int\biggl(\Bigl(\hat\vi_1 \frac {\d\Ga}{\d\vi_2}
    -Y_2q_1 +\hat\vi_2\frac {\d\Ga}{\d\vi_1} -Y_1q_2\Bigr)\cr
  & \phantom{\int }
+  s_{\Gacl}\Bigl(-\xi_A m\bar c\hat\vi_1+e\bar c\hat\vi_1^2-e
     \bar c\hat\vi_2^2\Bigr)\biggr)\cr
=\ue_5 & s_{\Gacl}\int\biggl(Y_2\hat\vi_1+Y_1\hat\vi_2 +\bar c
      (-\xi_Am\hat\vi_1+e\hat\vi_1^2-e\hat\vi^2_2)\biggr)\/.\cr}
\EQN\F333$$
The breaking is a variation with respect to the classical WI:
$$\eqalign{
(W_1\Ga)^{(\leq 1)}&=
\ue_5 s_{\Gacl}W\int(-Y_1\hat\vi_1+\bar c\hat\vi_1\hat\vi_2)\cr
&=\ue_5 W s_{\Gacl}  \int(-Y_1\hat\vi_1+\bar c\hat\vi_1\hat\vi_2)\cr}
\EQN\F335$$
Hence we are able to write the variation as a local counterterm
to $\Ga$, establishing thereby the deformed 1-loop WI:
$$
W_1\left(\Ga -\ue_5 s_\Ga\int(-Y_1\hat\vi_1+\bar c\hat\vi_1\hat\vi_2)
\right)=O(\hbar^2)\/.
\EQN\F336$$
It remains only to be seen that the counterterm $u_5s_\Ga\int(\ldots)$
can be added to $\Ga$ without spoiling the ST identity. But this
is clear because the counterterm is a BRS invariant whose coefficient
we can fix as we wish.

More explicitly it is the parameter $x_2$ \queq{\F337}, which is
equal to $x_1$  in the tree approximation \queq{\F317} and has
to be adjusted also in 1-loop order.
If we write
$$
x_2=x_1+x_2^{(1)}
\EQN\F338$$
we can determine $x^{(1)}_2$ as the solution of
$$\eqalign{
&\Bigl(\Ga_{Y_2q_2}-\Ga_{q_1Y_1}- \frac me (1+ \ue )\bigl(\hat \xi_A
\Ga _{\vi_2Y_1q_2}-
\xi_A \Ga_{{\hat\vi}_2Y_1q_2}\bigr)
\Bigr)_{\big|_{p^2=\kappa^2}} = 0\cr
}
\EQN\F339$$
i.e.\ in 1-loop (c.f.~\queq\F318 )
$$
x_2^{(1)}= - \frac me \Bigl((1-x\xi_A)
 \Ga_{\vi_2Y_1q_2}-\xi_A \Ga_{\hat\vi_2Y_1q_2}
\Bigr)_{\big|p^2=\kappa^2}
\EQN\F340 $$
Then the rigid WI holds  at the one-loop order.

In higher orders one proceeds by induction. One starts by assuming
 that a rigid Ward-identity of the following form is valid to n-loop
order
$$
\Bigl( W_n \Ga \Bigr) ^{(\le n)} = 0
\EQN\E341 $$
where $W_n $ is defined by the sum of the classical WI operator and
higher orders deformations analogously to \queq{\F330}:
$$W_n = W + \sum  _{i=1} ^n \d W ^{(i)}
\EQN\E342 $$
{}From there  one concludes by the same reasoning as in 1-loop order
that a deformed WI also holds at order $n+1$, if one adjusts
the parameters $\hat \xi_A $ and $x_2$ order by order appropriately.

The final form of the WI valid to all orders
$$\eqalignno{
 W_{\infty} \Ga  = &\, 0  &\EQ\E340A\cr
W_{\infty}
\equiv &\int\biggl(
-\vi_2\frac {\d }{\d\vi_1}
+(1+u)(\vi_1 -\hat \xi _A \frac me )
\frac {\d }{\d\vi_2}
 +  Y_1 \frac {\d }{\d Y_2} -(1+u) Y_2\frac{\d }{\d Y_1} & \cr
  & \phantom{\int} \, - \hat\vi_2\frac {\d }{\d\hat\vi_1} +
(1+u)(\hat\vi_1-\xi_A \frac me )
\frac{\d }{\d\hat\vi_2}
 - q_2\frac {\d}{\d q_1} +(1+u)q_1\frac{\d}{\d q_2}
\biggr)&
\subeqbegin \cr}
$$
can be immediately compared with the general deformed WI operator
$W^{\rm gen}$
\queq{\F316} of the classical approximation:
 We see that the deformation is
indeed as the general classical solution suggested.
$$W_{\infty} = z W^{\rm gen}
\EQN\E340B
$$
 Multiplying
\queq{\F316} with $z = \sqrt{z_1/z_2}$ we can identify the deformation
as
$$
{z_1\over z_2}={1+\hat z_1\over 1+\hat z_2}
                =1+ u
\EQN\F341$$
i.e. in 1-loop order
$$\hat z^{(1)}_1-\hat z^{(1)}_2=\ue
\EQN\F342$$
This combination of the wave function renormalizations is thus
independent of how one removes the infinities, but depends only
on the prescribed normalization.

 \chap{The local Ward-identity}

In the {\it abelian} Higgs model one can construct from the
deformed rigid WI \queq{\E340A} a local WI, which expresses
invariance of the Green functions under deformed local
gauge transformations. In contrast to QED this gauge invariance
does not characterize the spontaneously broken
model, but has to be derived from
the Slavnov-Taylor identity.

 We define the local WI operator from the rigid one, \queq{\F316} and
\queq{\E340A, \shorttag \E340B}, by taking
away the integration:
$$\eqalignno{
&W^{\rm gen}  \equiv \int dx {\rm w}^{\rm gen}(x) &\EQ\F41 \cr
{\rm w}^{\rm gen} (x) \equiv & \cr
-\zze& \vi_2{\d\over\d\vi_1}+\zez (\vi_1-\hat \xi_A \frac me )
                        {\d\over\d\vi_2}
-\zez Y_2{\d\over\d Y_1}+\zze Y_1{\d\over\d Y_2}\cr
-\zze&\hat\vi_2{\d\over\d\hat\vi_1}+\zez
\Bigl(\hat\vi_1-\xi_A \frac me   \Bigr)
{\d\over\d\hat\vi_2}-\zze q_2 {\d\over\d q_1}+\zez q_1{\d\over\d q_2}&
\subeqbegin\cr}
$$

In analogy to the treatment of the rigid WI we first
want to study the application of the local
 WI operator  on the general classical action
  \queq{\F38}. One verifies immediately
 $$
\Bigl( \hat e\sqrt{z_1z_2} {\rm w}^{\rm gen}(x) -\p{\d\over\d A}
\Bigr)\Gacl
^{\rm gen} =\dalam B\/.
\EQN\F41neu$$
Using the normalization conditions we find in the tree approximation
$$
\Bigl( e {\rm w}(x) -\p{\d\over\d A} \Bigr)\Gacl =\dalam B\/.
\EQN\F41tree $$
where ${\rm w}(x)$ ist the original undeformed local WI operator
(cf.~\queq{\F322A}).

In order to proceed to higher orders we have again  to classify the
operators according to their transformation properties with respect to
BRS and according to their quantum numbers.
{}From the WI \queq{\E340A}
$$
W^{\rm gen}\Ga =0
\EQN\F43$$
it follows with the help of the action principle, that
$$
{\rm w}^{\rm gen}(x)\Ga =\(\p^\mu j_\mu\)_4\cdot\Ga
\EQN\F44$$
where the insertion $\p^\mu j_\mu$ is a total derivative which has
dimension four, $\phi\pi$-charge zero and is odd under charge
conjugation.
Like $W^{\rm gen}$ the local ${\rm w}^{\rm gen}(x)$ is BRS symmetric and
$\p^\mu j_\mu$  is therefore a BRS invariant (cf.~\queq{\F322 a, b})
$$
0={\rm w}^{\rm gen}(x) s(\Ga) = s_\Ga \Bigl(\(\p j\)\cdot\Ga\Bigr)
\EQN\F45$$
The same characterization is true for the operator
$\pmo {\d\Ga\over\d A^\mu}$.
This is most easily seen by differentiating the ST identity
 \queq{\F37}
$$
s(\Ga)\equiv\int\p c {\d\Ga\over\d A}+B{\d\Ga\over\d\bar c}+
{\d\Ga\over\d\underline{Y}}\cdot {\d\Ga\over\d\uvi}+ \underline{q}
{\d\Ga\over\d\hat{\uvi}}=0
\EQN\F411$$
with respect to $c$. From there we obtain
$$
-\p{\d\Ga\over \d A}
-s_\Ga \Bigl({\d\Ga\over\d c}\Bigr)=0\/,
\EQN\F412$$
i.e.~it is not only a BRS invariant, but moreover a BRS variation.
Proceeding now order by order we get in 1-loop
$$ \Bigl(e{\rm w}^{\rm gen}(x)
 \Ga -\p{\d\Ga\over \d A}\Bigr)^{(\le 1)} = \p^\mu j_\mu^{(1)}
\EQN\E41
$$
Following the discussion above a short calculation shows, that
 a basis for $\p^\mu j_\mu^{(1)}  $ is
given by the two terms $\dalam B$
and $\p^ \mu j^{{\rm matter}}_\mu$, where we define
 (recall $\bar \vi _i = \vi_i - x \hat \vi_i $)
$$\eqalign{
j^{{\rm matter}}_\mu =\, & \bar\vi_2\pmu\bar\vi_1-(\bar\vi_1+
\frac me )\pmu\bar\vi_2\cr
        &+e A_\mu (\bar\vi_2^2+ (\bar\vi_1+\frac me )^2) \cr}
\EQN\F49$$
This field polynomial is indeed invariant under $s_{\Gacl }.$
One of the two basis elements can be replaced by the operator
$\p{\d\Gacl\over\d A} $ or $ {\rm w}(x) \Gacl$ respectively:
$$\eqalign{
\p{\d\Gacl\over\d A} &=  e {\rm w}(x) \Gacl -\dalam B \cr
             &= e \pmo j_\mu^{{\rm matter}}-\dalam B }
\EQN\E413
$$
Therefore we rewrite \queq{\E41} into the following form:
$$
\Bigl(e {\rm w}^{\rm gen}(x) \Ga -\p{\d\Ga\over \d A}\Bigr)^{(\le 1)} =
     a^{(1)} {\rm w}(x) \Gacl  + \dalam B
\EQN\E414
$$
The coefficient of $ \dalam B$ can be determined by testing on the
gauge fixing condition.
Shifting the variation $a^{(1)} {\rm w}(x) \Gacl  $ from the r.h.\ to
the l.h.s.\ one
gets the 1-loop local WI. Proceeding in the same way
by induction
 we derive to all orders  in perturbation theory
a local WI, which involves the deformed operator $ {\rm w}^{\rm gen}$
$$ \Bigl( e ( 1+a) {\rm w} ^{\rm gen} (x)
               - \p{\d \over \d A}\Bigr) \Ga = \dalam B
\EQN\415
$$
($a=O(\hbar)$).
When written in terms of $\Gamma$, the WI is most useful for the
purposes of renormalization, but for other applications its
formulation on $Z$, the generating functional for general Green
functions, is also interesting. We therefore present this version too.
$$\eqalignno{
 \Bigl( e ( 1+a)&{\rm w}^{\rm gen}\(J\) (x)
   + i\p^\mu J_\mu\Bigr) Z\(J\) = \dalam{\p Z \over \p J_B}
                                       &\EQ\F416\cr
{\rm w}^{\rm gen}\(J\) \equiv &&\subeqbegin\cr
\zze& J_{\vi_1}{\d\over\d J_{\vi _2}}-\zez J_{\vi _2}(-i\hat \xi_A \frac me +
                        {\d\over\d J_{\vi_1}})
-\zez Y_2{\d\over\d Y_1}+\zze Y_1{\d\over\d Y_2}\cr
-\zze&\hat\vi_2{\d\over\d\hat\vi_1}+\zez
\Bigl(\hat\vi_1-\xi_A \frac me   \Bigr)
{\d\over\d\hat\vi_2}-\zze q_2 {\d\over\d q_1}+\zez q_1{\d\over\d q_2}&
\cr}
$$

It is important to note that this WI does not characterize the
theory since it says nothing about the behaviour of the
$\phi\pi$-ghosts and does not permit to conclude that $\p A$
is a free field. The inhomogeneous contributions in ${\rm w}(x)$
prohibit this conclusion. I.e.\ unlike the unbroken case one
{\it has\/} to use the ST identity for the proof of unitarity
and for a scheme independent characterization of the model.
The local WI on the other hand permits one to study the fate of
$\p A$ when inserted in Green functions as being closely related
to the divergence of the current.

 \chap{The Callan-Symanzik equation}

The Callan-Symanzik (CS) equation describes the response of the
system to scaling of all independent
parameters carrying dimension of mass.
In the present  context of rigid invariance this
is of special interest, because
consequences of the underlying
symmetry manifest themselves most clearly as
relations of different coefficient functions.

\sect{The classical approximation}

In the classical approximation scale invariance is broken
$$
\smdm
\Gacl\equiv (m\p_m+m_H\p_{m_H}+\kappa\p_\kappa)\Gacl =\Delta_m
\EQN\F51$$
by all terms in the classical action with dimension less than
or equal to three.
We can immediately calculate $\Delta_m$ and find:
$$\eqalign{
\Delta _ m =&\, m^2 A^2 - m_H ^2 \bar \vi ^2 _1 \cr
            &\, - m \p \bar\vi _2 A  + e m \bar \vi _1 A^2
                 - \frac 12 \frac {m_H^2}{m} e
                 \bar\vi_1(\bar \vi_1^2+ \bar \vi_2 ^2) \cr
            &\, + m
  s\bigl(\bar c   (\hat \xi_A \vi _2 - \xi_A \hat \vi _2)\bigr)   }
\EQN\E51
$$
where $\bar \vi_i = \vi _i - x \hat \vi_ i $ \queq\F39 \
 and
$\hat \xi _A = - 1 + x \xi_A $ in the tree approximation
\queq{\F318}.
It is obvious that $\Delta _m $ is
 even under charge
conjugation. As a consequence of the BRS invariance of the theory
it is in addition BRS invariant as can be seen more abstractly by
applying $\smdm$ to the ST identity \queq{\F37}
$$
0=\smdm s(\Gacl)=s_{\Gacl}(\smdm\Gacl)=s_{\Gacl}(\Delta_m)
\EQN\F52$$
with $s_\Ga$ given in \queq{\F322 b}.
Furthermore
 it has a certain covariance with respect to the WI operator of
rigid symmetry $W$\queq\F322A .
$$ \( W , \smdm  \) \Gacl  = W \Delta _m = \hat \xi_A  \frac me
\int{\d \Gacl\over \d \vi_2} +  \xi _ A  \frac me
\int{\d \Gacl\over \d \hat \vi_2}
\EQN\E52
$$
 We have
to  check now, whether
$\Delta_m$ is uniquely determined by these characteristica.
Because, if so, we can proceed this way to all orders, where the
evaluation of $\Delta_m$ is not possible with explicit coefficients.
This analysis also gives
  a complete classification of the 2 and 3 dimensional
insertions appearing in the breaking of scale invariance.

The  BRS invariant terms contributing
 to $\Delta_m$ are quickly listed. There is
one invariant which is not a variation, the other terms are variations:
$$
\int (\bar\vi_1^2+2 v\bar\vi_1+\bar\vi_2^2)\/,
    s_{\Gacl}\int Y_1\/,\/s_{\Gacl}\int\bar c\vi_2,
      s_{\Gacl}\int\bar c\hat\vi_2
\EQN\EINS
$$
Since in higher orders it is preferable to deal with differential
operators instead of insertions, we replace
$s_{\Gacl}\int\bar c\vi_2$ by $\int{\d \Gacl \over \d \hat\vi_1}$
which according to
$$
{\d\Gacl\over\d\hat\vi_1}=x{\d\Gacl\over\d\vi_1}-x e~s_{\Gacl}
(\bar c\hat\vi_2)-e~s_{\Gacl}(\bar c\vi_2)
\EQN\F53$$
is also a variation. Hence $\Delta_m$ can be represented in the
form
$$\eqalign{
\Delta_m
         =&u_{inv}\int(\bar\vi_1^2+2\bar v\bar\vi_1+\bar\vi_2^2)+\int
        \Bigl(u_1{\d\Gacl\over\d\vi_1}+u_2{\d\Gacl\over\d\hat\vi_1}
        +u_3(B\hat\vi_2-\bar c q_2)\Bigr)\cr}
\EQN\F54$$
Testing first on the gauge condition with respect to $B$ \queq{\F312}
we find
$$
u_3+eu_1 =-\hat\xi_A m\qquad u_2=-\xi_A{m\over e}
\EQN\F56$$
or
$$\eqalign{
\smdm \Gacl = & u_1\int{\d\Gacl\over\d\vi_1}-\xi_A {m\over e}\int
{\d\Gacl\over\d\hat\vi_1}-(\hat\xi_Am+eu_1)\int(B\hat\vi_2-\bar c q_2)\cr
 &\qquad +u_{inv}\int (\bar\vi^2_1+2 v\bar\vi_1+\bar\vi_2^2)\cr}
\EQN\F57$$
So far we can get with BRS invariance alone.
But even without all external
fields the decomposition into the variation $s_{\Gacl}Y_1$ (first term)
and the non-variation (last term) would not be unique. At this point
rigid invariance has to be used:
 $\Gacl$ satisfies the rigid WI \queq{\F322A}  and
according to \queq{\E52} we extend $\smdm$ to $\tsmdm$ which by definition
commutes with $W$:
$$
\tsmdm \equiv \smdm +\hat \xi_A{m\over e}
 \int {\d\over \d\vi_1}+\xi_A {m\over e}\int {\d\over
\d\hat\vi_1} \qquad \(W,\tsmdm \)= 0 \/.
\EQN\F58$$
Hence we identify
$$
u_1 = -\hat \xi _A \frac me
\EQN\F59$$
and rewrite \queq\F57 \ into the symmetric form
$$\tsmdm \Gacl =u_{inv}\int(\bar\vi_1^2+2 v\bar \vi_1+\bar\vi_2^2)
\EQN\F510$$
The coefficient $u_{inv}$ cannot be determined by symmetry considerations;
one can calculate it by
testing with respect to $\vi_1$
$$
u_{inv}=\frac 12 m^2_H
\EQN\F511$$
Introducing a further external field $\hat\vi_0$ of dimension 2,
even under charge conjugation,
invariant under BRS  and rigid transformations
coupled to the invariant $\bar\vi_1^2 +2 v\bar\vi_1+\bar\vi_2^2$
we can
finally write the CS equ.\ in the classical approximation, where
$  \hat\xi_A=-1 + x \xi_A, $ as
$$
\smdm\Gacl ={m\over e}(1 -x \xi_A) \int {\d\Gacl\over\d\vi_1}-\xi_A{m\over e}
\int {\d\Gacl\over\d\hat\vi_1}+\ha m^2_H\int {\d\Gacl\over\d\hat\vi_0}
\EQN\F512$$
One verifies immediately that \queq{\F512} conincides with the explicit
determination of $\Delta _ m$ \queq{\E51}.

Summarizing the result of these considerations in the tree
approximation we can state that the r.h.s.\ of the CS equ.\ \queq{\F512}
is unique once we require invariance under BRS, rigid transformations
and charge conjugation and limit its dimension by three.

\sect{Higher orders}

We shall try to follow closely the reasoning of the tree
approximation. The action principle tells us that
$$
\smdm \Ga =\Delta_m\cdot\Ga
\EQN\F513$$
where $\Delta_m$ is now an insertion of power counting four,
even under charge conjugation and still BRS invariant due to
$$
0=\smdm s(\Ga)=s_\Ga (\smdm \Ga)=s_\Ga(\Delta_m\cdot \Ga )\/.
\EQN\F514$$
Hence we have to extend the above list of BRS invariant
 insertions \queq{\EINS , \shorttag \F53} by those of
dimension four, which we immediately give in the form of BRS-symmetric
operators. For the following operators the correspondence to
the BRS-symmetric insertions is again obvious
$$\eqaligntag{
s_\Ga\!\int \!\vi_1Y_1=\int \bigl(\vi_1{\d  \over \d {\vi_1}}
-Y_1{\d \over \d {Y_1}}\bigr)\Ga \qquad&
s_\Ga\!\int\! \hat\vi_1Y_1=\int
\bigl(\hat\vi_1 {\d\Ga \over \d {\vi_1}} -Y_1 q_1\bigr)\cr
s_\Ga\!\int\! \vi_2Y_2=\int \bigl(\vi_2{\d \over \d{\vi_2}}-
Y_2{\d \over \d {Y_2}}\bigr)\Ga\qquad&
s_\Ga\!\int\! \hat\vi_2Y_2=\int \bigl(\hat\vi_2{\d \Ga \over  \d {\vi_2}
}-Y_2 q_2)
&\EQADV\F515\subeqbegin\cr}
$$
whereas it requires a short calculation  to prove the independence
of the further operators, when acting on $\Ga$:
$$\eqaligntag{
&\int\bigl(A {\d \over \d A} + c {\d \over \d c}\bigr),\/
\int\bigl(B {\d \over \d B}+\bar c{\d \over \d {\bar c}}\bigr),\/
\int\bigl(\hat\vi_i
{\d \over \d{\hat\vi_i}}+q_i{\d\over \d {q_i}}\bigr)
\, \Ga
&\subeq\cr
&m_H\p_{m_H},\/e\p _e,\xi\p_\xi\, \Ga \/. &\subeq\cr}
$$
The insertion
$$\eqaligntag{
&\int ( B \hat \vi_1 \hat\vi_2 - \bar c q _1 \hat \vi_2 - \bar c \hat
\vi _1 q _2 )
&\subeq\cr}
$$
cannot be replaced by a BRS symmetric operator. Together with
\queq{\EINS} the insertions defined by \queq{\F515} constitute
a complete basis of BRS invariant insertions building up
$\Delta _m$. As one can easily convince oneself, the BRS symmetric
operator $\p _x \Ga $ is not independent, because
$$\eqalign{
\p _x
 \Gacl = \, - \int \Bigl(&\hat \vi_1 {\d \Gacl\over \d \vi _1 }
  + \hat \vi_2{\d \Gacl\over \d \vi _2 } + Y_1 q_1 + Y_2 q_2 \cr
- & e \xi_A m
s_{\Gacl}(\bar c \hat \vi_2)\Bigr) \quad \hbox{if}\quad x \ne 0}
\EQN\E516
$$
 The limit of $x= 0$, which is an allowed normalization,
 is not appropriately  represented by $\p_x \Ga$,
 and therefore we will remain with \queq{\F515}.

The tree approximation taught us the lesson that we should make
use of the rigid invariance if we want to construct a unique
r.h.s.\ of the CS equ.
Similarly to \queq{\F514} the insertion $\Delta _m$ also has
a certain covariance with respect to $W^{\rm gen}$ \queq{\E340A ,
\shorttag \E340B}
$$ \( W^{\rm gen} ,\smdm \) =
                    W^{\rm gen} \bigl( \Delta _m \cdot \Ga \bigr)
= \sqrt{ \frac {z_1 }{z_2}  }\frac me  \int\bigl(\hat \xi_A
{\d \Ga\over \d \vi_2} +  \xi _ A
{\d \Ga\over \d \hat \vi_2}  \bigl)
\EQN{\E54}
$$
which is the generalization of \queq{\E52} to higher orders.
Hence according to the derivation above we will consider instead
of $\smdm $ the symmetric version $ \tsmdm $ defined in
\queq{\F58}, which can be shown to commute with $W^{\rm gen}$, too, and
the insertion $\tilde\Delta_m\cdot\Ga$
$$
\tsmdm \Ga =\tilde\Delta_m\cdot\Ga
\EQN\F520$$
is  $W ^{\rm gen}$ symmetric.
Therefore we symmetrize first of all the
operators listed in \queq{\F515} with respect to the general rigid
WI operator $W^{\rm gen} $ \queq{\E340A}.
 The leg counting operators amongst those of
\queq{\F515} are  easily symmetrized. They read
$$\eqalign{
\N_s \equiv N_s - \hat\xi _A\frac me \int \!\d_{\vi_1}&
\equiv\int\Bigl((\vi_1 -\hat \xi_A \frac me )\d_{\vi_1}+\vi_2\d_{\vi_2}-
Y_1\d_{Y_1}-Y_2\d_{Y_2}\Bigr)\cr
\hat\N_s\equiv \hat N_s - \xi_A \frac me \int \!\d _{\hat\vi_1}&
\equiv\int\Bigl((\hat\vi_1-\xi_A\frac me )\d_{\hat\vi_1}+
\hat\vi_2\d_{\hat\vi_2}+q_1\d_{q_1}+q_2\d_{q_2}\Bigr)\cr
N_A &\equiv\int\Bigr(A\d_A+c\d_c\Bigr)\cr
                         N_{\scriptscriptstyle B}
 &\equiv\int\Bigl(B\d_B+\bar c\d_{\bar c}\Bigr)\cr}
\EQN\F517$$
The mixed operators $\hat \vi _i {\d \over \d \vi _i}$ are symmetrized
like the leg counting operators
$$\bar \N_s\equiv\bar N_s - \xi _A \frac me \int \! \d_{ \vi _1}
\equiv\int\Bigl((\hat\vi_1-\xi_A\frac me )\d_{\vi_1}+
\hat\vi_2\d_{\vi_2}\Bigr)
\EQN\E517$$
and the insertion
$$
\bar \N_s \Ga
+ \int \! (q_1Y_1+q_2Y_2)
\EQN\E518$$
is BRS and $W^{\rm gen}$ invariant.
Slightly more involved is the symmetrization of the differential
operators in \queq{\F515 c} $ \nabla  = m_H\p_{m_H}, e\p_e, \xi\p_\xi$.
Their symmetrized extensions $\tilde \nabla$   have the explicit form
$$\eqalign{
\tilde\nabla\equiv \nabla &+\int \Bigl( \nabla  (\hat \xi_A \frac me  )
\d_{\vi_1}+\nabla
    (\xi_A \frac me )\d_{\hat\vi_1} \Bigr)\cr
&-\frac 12 \nabla(\ln \frac {z_1}{ z_2} )\int
\Bigl(
(\vi_1 - \hat \xi_A \frac me )\d_{\vi_1}-Y_1 \d_{Y_1}\cr
&\phantom{-\frac 12 \nabla(\ln \frac {z_1}{ z_2} )\int}
 +(\hat\vi_1-\xi_A \frac me )
\d_{\hat\vi_1}+q_1\d_{q_1}\Bigr)\cr}
\EQN\F519$$
The insertion \queq{\F515 d} is not  invariant under rigid transformations
and can
therefore not contribute on the r.h.s. of \queq{\F520}.
When inspecting the operators of dimension three (compare \queq{\F54}
and \queq{\F512}) we find that only $\int{\d \over \d {\hat\vi_0}}$ can
contribute $W$-symmetrically.

 The basis of BRS and rigidly invariant
differential operators which are charge conjugation invariant and
have dimension less than or equal to four is thus provided by
\queq{\F517,\shorttag \E518 ,
\shorttag \F519} and $\int\d_{\hat\vi_0}$. The insertion
$\tilde\Delta_m\cdot\Ga$  \queq{\F520}
can therefore be decomposed in the following way
$$\eqalign{
\tilde {\cal C}\Ga\equiv
\Bigl(&\tsmdm +\beta_e e\tilde\p_e +\beta_{m_H}
m_H\tilde\p_{m_H}+\beta_\xi\xi\tilde\p_\xi\cr
&-\ga_s\N_s-\hat\ga_s\hat{\N}_s-
\bar \ga_s\bar {\N}_s-\ga_{\scriptscriptstyle A}
  \N_A-\ga_B\N_B-\alpha_{\rm inv} \int\d_{\hat\vi_0}\Bigr)\Ga\cr
= \,& \bar\ga_s\int(q_1Y_1 +q_2Y_2)\cr}
\EQN\F521$$
This is the CS-equ.\ in manifestly $W$-symmetric form.
Its most important feature is the appearance of the $\beta$-function
$\beta_{m_H}$ which is a consequence of the physical normalization
of the model. Since the scalar self-coupling $\lambda$ is then
not an independent parameter $\beta_{m_H}$ has to replace $\beta_\lambda$.
It is quite unclear, how in higher orders a formulation in terms of
unphysical parameters could be related to the one in physical
parameters (cf.\ \quref{\EK2,\EK3}).

Some additional information on the coefficients comes from testing on the
gauge condition \queq{\F312}:
$$
\eqaligntag{
\ga_B = &-\ga_A     &\EQADV\F523\subeqbegin\cr
\beta_\xi = &-2 \ga_A &\subeq\cr
\beta_e +\ga_A-\ga_s-\hat\ga_s = &\frac 12 (\beta_e e\p_e+\b_{m_H}m_H\p_{m_H}
+\b_\xi\xi \p_\xi    )
\ln \frac {z_1}{ z_2}
                    &\subeq\cr }
%
$$
One has to note, that
the last relation constitutes a connection of the anomalous dimension
of the external fields with the coefficient functions of the propagating
fields, but there is no similar relation for $\bar \ga _s$.

One further relation emerges from the validity of the local
WI \queq{\415}
$$ \Bigl( e \hat a {\rm w} ^{\rm gen} (x)  - \p{\d \over \d A}\Bigr) \Ga =
 \dalam B  \qquad \hbox{with} \quad \hat a = 1 + a
\EQN\415
$$
We  calculate the commutators
$$ \eqalign{
\(\tilde {\cal C}, e \hat a {\rm w} ^{\rm gen} (x) \) =&\, e
\Bigl( \beta _e \hat a +  (\beta_e e \p_e + \beta _{m_H} {m_H} \p_ {m_H} +
\beta _\xi \xi \p _\xi )
\hat a \Bigr) {\rm w} ^{\rm gen} (x) \cr
\(\tilde {\cal C}, \p{\d \over \d A}\)   = &\, \ga_A \p{\d \over \d A} \cr}
\EQN\F527$$
and with \queq{\F523 a}
$$
\tilde {\cal C } \,\dalam B =   \ga_A \dalam B
\EQN\F527E
$$
Combining \queq{\F527} and \queq{\shorttag \F527E} with the local
WI \queq{\415}
$$
\Bigl\(\tilde {\cal C}, e \hat a {\rm w} ^{\rm gen} (x)
- \p{\d \over \d A} \Bigr \) \Ga = \tilde {\cal C} \, \dalam B
\EQN\E528
$$
we get the relation:
$$
\ga _A = \beta_e + (\beta_e e\p_e + \beta _{m_H}{m_H} \p_ {m_H} -
2 \ga _A \xi \p _\xi )\ln (1+a)
\EQN\E529
$$
Here we have also inserted equ.\ \queq{\F523 b}.
Since $a $ is  of order $\hbar$,
the second term does not contribute in one-loop, so in
one-loop we have
$$
\ga^{(1)}_A=\b_e^{(1)}\, ,
\EQN\F532 $$
a relation which is also
 well-known from the unbroken version of the model.
Higher orders are then recursively determined.

In \queq{\F521} we have
 given the CS-equation in a manifestly $W$-symmetric
form using the symmetrized operator $\N_I $ and $\tilde \nabla .$
For calculations, as for example the derivation of the leading logarithm
behaviour, it is much more convenient to rewrite it into the
usual form, which separates the hard and soft breaking on the
left and right hand side of the CS-equation.
Summarizing thereby also the relations we have derived in
\queq{\F523} and \queq{\E529} we end up with the following form:
$$\eqalign{
  {\cal C} \Ga\equiv \Bigl(
 &\smdm +\beta_e e\p_e +\beta_{m_H} m_H\p_{m_H}
        -\ga_s N_s-\hat\ga_s\hat{N}_s- \bar \ga_s\bar {N}_s\cr
 &-\ga _1 \int \! ( \vi _1\frac {\d}{\d\vi_1} -Y_1\frac {\d}{\d Y_1}
       + \hat \vi _1\frac {\d}{\d\hat \vi_1} + q_1\frac {\d}{\d q_1} )
    -\ga_{\scriptscriptstyle A} ( N_A-N_B+ 2 \xi \p_\xi) \Bigr) \Ga \cr
 \phantom{{\cal C} \Ga} = \phantom{\Bigl(} &
                     - \frac me \Bigl( (\hat \xi _A  + \alpha _1) \int\!
                \frac {\d}{\d\vi_1} + ( \xi _A  + \hat \alpha _1) \int\!
                \frac {\d}{\d\hat \vi_1}  -\alpha_{\rm inv} \int \!
                \frac {\d}{\d\hat \vi_0} \Bigr)\Ga  \cr
 & + \bar\ga_s\int(q_1Y_1 +q_2Y_2)\cr  }
\EQN\F521
$$
with
$$
\eqalign{
\ga _1      =& \frac 12 (\beta_e e \p_e + \beta _{m_H} {m_H} \p_ {m_H}
        -2 \ga _A \xi \p _\xi ) \ln \frac {z_1}{z_2} = O(\hbar ^ 2)\cr
\hat \ga _s =& \beta _e + \ga _A -\ga _s + \ga _1 \cr
     \ga _A =& \beta_e + (\beta_e e\p_e + \beta _{m_H}{m_H} \p_ {m_H}
-2 \ga_A \xi \p _\xi )\ln (1+a) \cr
\alpha_1    =&  \Bigl((  \ga _s \hat \xi _A +
                        \ga _1 \hat \xi _A -\beta _e
                          \hat \xi _A+  \bar \ga _s \xi _A  )\cr
             &+(\beta_e e \p_e + \beta _{m_H} {m_H} \p_ {m_H} -
             2\ga_A  \xi \p _\xi ) \hat \xi_A  \Bigr) \cr
\hat \alpha_1 =&  \Bigl((   \hat \ga _s \xi _A +
                  \ga _1  \xi _A - \beta _e \xi _A )\cr
              &+ (\beta_e e \p_e + \beta _{m_H} {m_H} \p_ {m_H} -2
                  \ga _A  \xi \p _\xi )  \xi_A  \Bigr) \cr}
\SUBEQNBEGIN\ECSa
$$
The independent parameters are therefore the coefficient functions
$\beta_e, \beta_{m_H}, \ga _s$ and $\bar \ga _s $, which we give
in the one-loop order in the next section,
and the coefficient of the soft insertion $\a _{\rm inv}=
\ha e{m_H ^2 \over m} + O (\hbar)$.
 Whereas  in the one-loop order the hard
anomalies are independent of the normalization conditions we have
chosen, i.e.\ especially of $\frac {z_1}{z_2}  $, one immediately
verifies that one finds  corrections due to the deformation of
the rigid WI starting from two loop onwards.
Let us emphasize again that the identification of the soft terms
has been accomplished by use of the rigid symmetry.

\chap{Determination of coefficient functions}

In the preceding sections we have derived Ward identities and the CS
equ.\ in a way which is essentially independent of the scheme with
which one regularizes and renormalizes the theory.
To be complete and
to come closer to practical work
we determine the 1-loop coefficient functions of the CS-equation
 \queq{\F521}.
Furthermore we want to demonstrate that the deformation coefficient
of the Ward-identity is indeed non-trivial in the on-shell scheme.

\sect{Deformation coefficient of the rigid WI}

In sect.\ 3  we proved the WI \queq{\E340A,\shorttag \E340B} in which
the coefficient
$z_1/z_2$  indicates the possible deformation of the classical
approximation due to the normalization conditions \queq{\F219 d,e}.
$$\eqalign{
& W^{\rm gen}\Ga\cr
& \equiv \int\biggl(-\sqrt{\frac {z_2}{ z_1} }
\vi_2{\d\over\d\vi_1}
+\sqrt{\frac {z_1}{ z_2} }(\vi_1 - \hat \xi_A \frac me ){\d\over\d\vi_2}-
\sqrt{\frac {z_1}{ z_2} } Y_2{\d\over\d Y_1}+\sqrt{\frac {z_2}{ z_1} }
   Y_1{\d\over\d Y_2}\cr
& \phantom{\equiv \int \,}
 -\sqrt{\frac {z_2}{ z_1} }\hat\vi_2 {\d\over\d\hat\vi_1}+
   \sqrt{\frac {z_1}{ z_2} }(\hat\vi_1-\xi_A \frac me  )
   {\d\over\d\hat\vi_2}-\sqrt{\frac {z_2}{ z_1} }q_2{\d\over\d q_1}
   +\sqrt{\frac {z_1}{z_2} } q_1{\d\over\d q_2}\biggr)\Ga =0\cr}
\EQN\F316$$
We want to show in the following,
 that this coefficient is indeed non-trivial, i.e.\
${z_1\over z_2} \ne 1$,
and gets higher order corrections unless one chooses some special
unphysical normalization conditions.

The factor ${z_1\over z_2}
$ can be determined by testing the WI with respect
to $\vi_1$ and $\vi_2 $, it yields in momentum space
$$\eqalign{
&-\sqrt{ \frac {z_2}{z_1}  }
  \Ga_{\vi_1\vi_1}(p^2) + \sqrt{ \frac {z_1}{z_2} }
\Ga _{\vi_2\vi_2}(p^2) \cr & -
\sqrt{ \frac {z_1}{ z_2} } \frac me \bigl( \hat \xi_A
\Ga_{\vi_1\vi_2\vi_2}(p,-p,0) +\xi_A
\Ga_{\vi_1\vi_2\hat\vi_2}(p,-p,0) \bigr)= 0}
\EQN\F63$$
In order to project out the residua we
differentiate  with respect to the momentum $p^2$ and
 get the following equation for the 1-loop coefficient
$\ue$ defined by ${z_1 \over z_2} = 1 + u$ \queq{\F341}:
$$\eqalign{
& \frac 12 \ue -\p_{p^2}
\Ga_{\vi_1\vi_1}^{(1)}(p^2)
+ \frac 12 \ue +\p_{p^2}\Ga_{\vi_2\vi_2}^{(1)}(p^2) \cr
+& \frac me  \p_{p^2}
\Bigl( (1-x \xi_A) \Ga_{\vi_1\vi_2\vi_2}^{(1)}(p^2)-\xi_A
\Ga^{(1)}_{\vi_1\vi_2\hat\vi_2}(p^2) \Bigr) = 0}
\EQN\F64$$
To determine $u^{(1)}$ explicitly one had to calculate the
2-point and 3-point function according to \queq{\F64}, but
we shall now show
by simple arguments that $u^{(1)}$ depends only on
 the wave function normalization and turns out to differ from zero
within physical normalization schemes.

For this purpose we use the normalization conditions \queq{\F219 d,e}
in a slightly generalized form:
$$\eqalign{
\p_{p^2}\Ga_{\vi_2\vi_2}
 (p^2=\kappa^2)=& 1 \cr
\p_{p^2}\Ga_{\vi_1\vi_1}(p^2=\mu ^2)=&1
} \EQN\E65
$$
where $\mu$ is a further independent normalization point.
According to \queq\E65 \ we can rewrite $\p_{p^2}\Ga_{\vi_i\vi_i}$
into
$$ \eqalign{
\p_{p^2}\Ga_{\vi_1\vi_1 } =
                  & 1 + \p _{p^2} \Sigma_1 ^{(1)} (p^2, m^2, m_H^2)
    -\p _{p^2} \Sigma_1 ^{(1)} (p^2, m^2, m_H^2)\Big| _ {p^2 = \mu^2} +
 O (\hbar^2)
 \cr
\p_{p^2}\Ga_{\vi_2\vi_2 } =
            & 1 + \p _{p^2} \Sigma_2 ^{(1)} (p^2, m^2, m_H^2)
    -\p _{p^2} \Sigma_2 ^{(1)} (p^2, m^2, m_H^2)\Big| _ {p^2 = \kappa^2}
+ O (\hbar^2) \cr}
\EQN\E66
$$
$\Sigma_i ^{(1)} (p^2, m^2, m_H^2)$ is the usual self energy
calculated in a specific scheme as e.g.\ in the $\widebar {MS} $-scheme.

Since  equ.\ \queq{\F64} is true for all momenta $p^2$, the momentum
dependence has to cancel, and it can therefore be evaluated at every
 convenient
value. One possiblity is at a momentum large compared to the masses:
There the three-point functions vanish asymptotically
and according to the CS-equation one finds for the self energy
contribution (see~$(6.10)$)
$$\lim_{\scriptscriptstyle p^2 \to -\infty}
\p _{p^2} \Sigma_1 ^{(1)} (p^2, m^2, m_H^2) =
\lim_{\scriptscriptstyle p^2 \to -\infty}
\p _{p^2} \Sigma_2 ^{(1)} (p^2, m^2, m_H^2) \rightarrow
           - \ga_s ^{(1)} \ln\frac {p ^2}{m^2} \/,
\EQN\E67
$$
where we have  normalized $\p_{p^2} \Sigma   _i$ appropriately
in the asymptotics because it is determined only up to constants.
Therefore, in order to determine $\ue$ one remains with
$$
\ue + \p _{p^2} \Sigma_1 ^{(1)} (p^2, m^2, m_H^2)\Big| _ {p^2 = \mu^2}
 -\p _{p^2} \Sigma_2 ^{(1)}
        (p^2, m^2, m_H^2)\Big| _ {p^2 = \kappa ^2} = 0
\EQN\E68
$$
Inspection of the diagrams shows that $ \Sigma_1 ^{(1)}$  and $ \Sigma_2
 ^{(1)}$ differ at least by contributions built up from diagrams with
 trilinear  $\vi$-vertices.
 To be more specific
we fix the residuum of the unphysical particle $\vi_2$ at a
momentum $\kappa$ large compared to the masses, i.e.
$$
   \p_{p^2}\Ga_{\vi_2\vi_2} (p^2=\kappa^2, |\kappa ^2 | \gg m^2 , m_H ^2)= 1
 \EQN\E69
$$
Such normalization conditions are e.g.\
 implicitly chosen if one calculates
the self energy of $\vi_2$ in the $\widebar {MS} $-scheme.
With \queq{\E69} equ.\ \queq\E68 \ simplifies to
$$
\ue +\p _{p^2} \Sigma_1 ^{(1)} (p^2, m^2, m_H^2)\Big| _ {p^2 = \mu^2}
 + \ga_s ^{(1)} \ln \frac {\kappa ^2}{m^2} = 0
\EQN\E70
 $$
Hence   $\ue  $ will differ from zero, unless  we fix the wave function
renormalizations  {\it both} at the same
asymptotic momentum, $\mu^2 = \kappa^2 $. Such unphysical normalization
 conditions are not
useful if one wants to calculate the $S$-matrix, because there the wave
function normalization of the physical particle
at finite momentum is needed. We conclude
that otherwise, especially
within the physical on-shell scheme $\mu^2 = m_H^2$  \queq{\F219},
  the WI-operator is indeed
deformed by $\ue_1\not= 0$.

\sect{Coefficients of the CS-equation}

Due to the inclusion of the external field multiplet $\hat\vi$ and the
parametrization as arising from physical normalization conditions the
CS-equ.\ \queq{\F521} has a slightly unconventional form. We therefore
  calculate the independent 1-loop coefficients explicitly.

Let us first determine $\ga_s$. It is found by studying the action of
the CS operator $\C$ \queq\F521 \
 on $\p_{p^2}\Ga_{\vi_1\vi_1}$. Since the CS-equ.\ is valid
for all momenta we can go to infinite momentum, where the r.h.s.,
the soft insertions, vanish. Therefore we get
$$\eqalign{
&\smdm \p_{p^2}\Ga_{\vi_1\vi_1}^{(1)} -
2\ga_s ^{(1)} \p_{p^2}\Ga_{\vi_1\vi_1}^{(o)} = \cr
&\smdm \p_{p^2}\Ga_{\vi_1\vi_1}^{(1)} - 2 \ga_s ^{(1)} \, {
\buildrel
p^2 \to -\infty \over \longrightarrow }\, 0 }
\EQN\F65$$
Hence we are left with the
calculation of $\p_{p^2}\Ga_{\vi_1\vi_1}$ at asymptotic momentum.
This yields
$$
\ga^{(1)}_s =-{e^2\over 16\pi^2}(3-\xi)
\EQN\F67$$
Next we calculate $\beta^{(1)}_e$. According to \queq{\F532} this
simplifies to the calculation of $\ga_A^{(1)}$, given by the vector
self-energy.  For asymptotic momentum analogous arguments are valid for the
disappearance of three-point contributions and the evaluation of the
two-point
functions leads to
$$
\ga^{(1)}_A ={1\over 16\pi^2}\cdot{1\over 3}
                 e^2={1\over e}\beta^{(1)}_e\/,
\EQN\F68$$
 For
 $\b_{m_H}$
the relevant vertex function is the four-point function
$\Ga_{\vi_1\vi_1\vi_1\vi_1}$:
$$
\smdm \Ga^{(1)}_{\vi_1\vi_1\vi_1\vi_1}-6 \b^{(1)}_ee^2 {m^2_H\over m^2}
-6 \b^{(1)}_{m_H}
e^2 {m^2_H\over m^2}+12\ga^{(1)}_s {m_H^2\over m^2}  \,
{ \buildrel p^2 \to -\infty \over \longrightarrow} \, 0
\EQN\F612$$
and one gets
$$
\b^{(1)}_{m_H} = {e^2 \over 16\pi^2} \Bigl( 5 {m_H^2 \over m^2} +
6 { m^2 \over m_H^2} - {19 \over 3} \Bigr)
\EQN\F613
$$
The functions $
 \ga_s^{(1)}, \beta_e^ {(1)} $ and $\beta_{m_H}^{(1)} $ are determined
by the vertex functions of the quantum fields and are in agreement with
the symmetric theory.

The coefficient function
$\bar \ga_S$ is determined by vertex functions including external fields.
It can be found by acting with $\C$
on $\p_{p^2}\Ga_{\vi_1\hat\vi_1}$. Again taking the limit of
infinite $p^2$ we find
$$
\smdm\p_{p^2}\Ga_{\vi_1\hat\vi_1}^{(1)}-\ga_s^{(1)} x-\bar\ga_s^{(1)}-
(2{\beta_e^{(1)}\over e}-\ga_s^{(1)}) x \,
 {\buildrel p^2 \to -\infty  \over \longrightarrow} \, 0
\EQN\F610$$
(we have used \queq{\F523 c}). The evaluation of the respective diagrams
leads eventually to the result
$$
\bar\ga_s^{(1)}={2\over 16\pi^2}e^2+2\Bigl(\ga_s^{(1)}-
{\b_e^{(1)}\over e}\Bigr) x
\EQN\F611$$
One has to note, that $\bar \ga_s $ depends on the parameter $x$
and is different from zero, even if we take $x=0$.

\chap{Discussion and conclusions}

In models with gauge invariance the rigid invariance encodes the
physical consequences of the symmetry: e.g.\ the conservation of quantum
numbers in physical processes as the consequence of conserved charges;
the arrangement of particles in multiplets; definite relations amongst
physical amplitudes. In case that the symmetry is not broken it can be
implemented easily in every renormalization scheme and requires
no special care. Whether (formal) unitarity is guaranteed by a
local WI (in the abelian case) or by the ST identity (in the
non-abelian case) does not really matter, the rigid WI can just
be written down naively. In the case of spontaneous breakdown of the
rigid symmetry the situation changes. A conserved charge does no longer
exist, the consequences of the symmetry reside entirely in relations
amongst amplitudes, which have to be deduced from a WI. Unitarity
(formal or full) can only be deduced from the ST identity which also
serves as the unique characterization of the model. The WI for the
rigid symmetry is no longer trivial to deduce, i.e.\ one has
to organize radiative corrections with quite some care. It has to
be formulated explicitly in accordance with the ST identity and its
form turns out to depend on the normalization conditions. In the
abelian Higgs model which we treated in the present paper we
have seen that higher orders deform the WI in a well specified way
once we stick to physical on-shell normalization conditions.
Hence the relations amongst Green functions found in lowest order
as a consequence of the WI are deformed very specifically in higher
orders. The ST identity guarantees unitarity and restricts the rigid
WI, but does not uniquely fix it. Once the rigid WI has been constructed
it is immediately useful for the formulation of the CS equ.
It would be quite difficult to handle there the soft terms
i.e.\ to define the higher orders without the rigid invariance.
In more complicated theories like the standard model it is virtually
impossible to proceed without it. There a deformation of the classical
WI operator not only involves relative factors for the fields within
one multiplet, but also for the several non-abelian generators
relative to each other, in particular the orientation of the
electromagnetic $U(1)$ relative to the remainder of the group.
The details will be reported elsewhere.

As an important technical tool --
 again indispensable in the standard model --
we have introduced a doublet of external scalar fields in order to
formulate the 't~Hooft gauge and its breaking of the
rigid symmetry in a controllable way. They render the rigid
WI homogeneous and thus manageable
in higher orders. How they form the building block to
a background field formulation of the model remains to be explored.
\bigskip

{\it Acknowledgement} \quad We are grateful to B.\ Kniehl for
 drawing our
attention to the references \quref{\BARBIERI, \DEGRASSI}.
\endpage

\centerline{\bf Appendix \quad The Propagators}

The bilinear part of the classical action reads
$$\eqalign{
\Ga_{{\rm bil.}} = &\int\Bigl(-\frac 14 (\pmu \An-\pnu \Am)
   (\pmo A^\nu -\p^\nu A^\mu)\cr
&\quad +\frac 12 (\p\vi_1\p\vi_1+\p\vi_2\p\vi_2)-m\p {\vi_2}A+
        \frac 12 m^2 A^2- \frac 12  m^2_H\vi^2_1\cr
&\quad +\frac 12 \xi B^2+B(\p A+\xi_A m\vi_2)
        -{\bar c}\, \dalam c-\xi_Am^2\bar c c \Bigr)\cr}$$
It gives rise to the following propagators:
$$\eqalign{
 G_{\vi_1\vi_1} (p,-p)&={i\over p^2-m^2_H}\cr
 G_{BB}(p,-p) &=0\cr
 G_{BA_\mu}(p,-p)&={-p^\mu\over p^2-\xi_Am^2}\cr
 G_{B\vi_2}(p,-p)&={-im\over p^2-\xi_A m^2}\cr
 G_{\vi_2\vi_2}(p,-p)&=i{p^2-\xi m^2\over (p^2-\xi_A m^2)^2} = i
\left({1\over p^2-\xi_A m^2}+{m^2(\xi_A-\xi)\over (p^2-\xi_Am^2)^2}
\right)\cr
G_{\vi_2 A_\mu}(p,-p) & = {-m(\xi -\xi_A) p _\mu \over (p^2-\xi_Am^2)^2}\cr
 G_{A_\mu A_\nu} (p,-p)&= \Bigl(\eta^{\mn}-{p^\mu p^\nu\over p^2}) G^T+
  {p^\mu p^\nu\over p^2} G^L\cr
 G^T (p^2) &={-i\over p^2-m^2}\cr
 G^L(p^2) &=i \left( {-\xi\over p^2-\xi_A m^2}
       +{(\xi_A-\xi)\xi_Am^2\over (p^2-\xi_A m^2)^2}\right)\cr}
$$
For the Fourier transformation we have used the conventions:
$$
 G_{\phi_a \phi _b} (x,y)= \int {d^4 p \over (2\pi )^4} G_{\phi_a \phi_b}
(p,-p) e^{ip(x-y)}
$$

\refout

\bye